\documentclass[10pt]{article}
\usepackage{graphicx}
\usepackage{amsmath}
\usepackage{amssymb}
\usepackage{caption2}
\begin{document}
\bibliographystyle{prsty}
\begin{center}
{\large {\bf \sc{  Analysis of  the hidden-charm pentaquark candidates in the $J/\psi \Sigma$ mass spectrum  via the  QCD sum rules }}} \\[2mm]
Zhi-Gang Wang \footnote{E-mail: zgwang@aliyun.com.  }, Yang Liu     \\
 Department of Physics, North China Electric Power University, Baoding 071003, P. R. China
\end{center}

\begin{abstract}
In this work, we adopt the diquark model, construct the diquark-diquark-antiquark  type  currents with the light quarks $uus$ in two octets, and  study the $uusc\bar{c}$ pentaquark states in the framework of   the QCD sum rules systematically.
We obtain the mass spectrum of the hidden-charm-singly-strange pentaquark states with the quantum numbers $IJ^{P}=1{\frac{1}{2}}^-$, $1{\frac{3}{2}}^-$ and $1{\frac{5}{2}}^-$. And we can search for those $P_{cs}$ states in the processes  $\Sigma_b^+\to P_{cs}^+\phi \to J/\psi \Sigma^+ \,\phi$ and $\Xi_b^0\to P_{cs}^+ K^- \to J/\psi \Sigma^+\, K^-$.
\end{abstract}



\section{Introduction}
Experimentally,  the  LHCb collaboration  has observed several pentaquark candidates, such as the $P_c(4380)$, $P_c(4450)$ \cite{LHCb-4380}, $P_c(4312)$, $P_c(4440)$, $P_c(4457)$ \cite{LHCb-Pc4312}, $P_{cs}(4459)$ \cite{LHCb-Pcs4459-2012}, $P_c(4337)$ \cite{LHCb-Pc4337}, $P_{cs}(4338)$  \cite{LHCb-Pcs4338}. On the other hand, the Belle and Belle-II collaborations observed the $\Upsilon(1{\rm S }, 2{\rm S})$ inclusive decays to the final states $J/\psi\Lambda$, and found  an evidence of the $P_{cs}(4459)$  with a local significance of $3.3\,\sigma$ \cite{Belle-Pcs4338-Pcs4459}.
Those pentaquark candidates always lie near the meson-baryon thresholds,
\begin{eqnarray}\label{Assign-MB}
\bar{D}\Sigma_c&:& P_c(4312) \, ,\nonumber \\
\bar{D}\Xi_c&:& P_{cs}(4338) \, ,\nonumber \\
\bar{D}^*\Xi_c &:& P_{cs}(4459) \, ,\nonumber \\
\bar{D}\Sigma^*_c&:& P_c(4380) \, ,\nonumber \\
\bar{D}^*\Sigma_c&:& P_c(4440/4457) \, ,
\end{eqnarray}
which favor the possible interpretations  as the hadronic molecular states \cite{Review-penta-mole-ZhSL-RPT,Review-penta-Esposito-RPT,Review-penta-Ali-PPNP,
Review-penta-mole-GuoFK-RMP,Review-penta-mole-LiuYR-PPNP,
Review-penta-mole-Brambilla-RPT,Review-mole-GuoFK-CTP,
Review-mole-WangB-PRT,Review-mole-GengLS-PRT,WangZG-Review}.
 But for the $P_c(4337)$,  it lies not far away from (not just in) the  $\bar{D}^* \Lambda_c$,
  $\bar{D} \Sigma_c$ and $\bar{D} \Sigma_c^*$ thresholds,  it is difficult to assign it as a molecular state without resorting to large coupled channel effects.

The QCD sum rules method  is a powerful theoretical tool in studying the exotic states, such as the tetraquark states, pentaquark states, molecular states, etc \cite{WangZG-Review,WangZG-landau-PRD,WangZG-IJMPA-Two-part,Nielsen-review}.
In the QCD sum rules for the hidden-charm (hidden-bottom) tetraquark and pentaquark (molecular) states, the integrals,
 \begin{eqnarray}
 \int_{4m_Q^2(\mu)}^{s_0} ds \,\rho_{QCD}(s,\mu)\exp\left(-\frac{s}{T^2} \right)\, ,
 \end{eqnarray}
are sensitive to the  energy scales $\mu$, where the $\rho_{QCD}(s,\mu)$ are the QCD spectral densities,  the $T^2$ are the Borel parameters, and the $s_0$ are the continuum threshold parameters.
In Ref.\cite{WangHuangtao-2014-PRD}, we tentatively identified the $X(3872)$ and $Z_c(3900)$ as   the diquark-antidiquark type  hidden-charm tetraquark states,  and  studied the energy scale dependence of the QCD sum rules for the exotic states ($X$, $Y$, $Z$) firstly. Subsequently,  we suggested an energy scale  formula,
\begin{eqnarray}
\mu&=&\sqrt{M^2_{X/Y/Z}-(2{\mathbb{M}}_Q)^2} \, ,
 \end{eqnarray}
 with the effective heavy quark  masses ${\mathbb{M}}_Q$ to determine the optimal energy scales, which work very well   \cite{Wang-tetra-formula,WangZG-mole-formula,WangHuang-2014-NPA}.
Then we took into account the light-flavor $SU(3)$ mass-breaking effects and modified the energy scale formula,
\begin{eqnarray}
\mu&=&\sqrt{M^2_{X/Y/Z}-(2{\mathbb{M}}_Q)^2}-\kappa {{\mathbb{M}}_s} \, ,
 \end{eqnarray}
by introducing an effective $s$-quark mass ${\mathbb{M}}_s$, where the $\kappa$ is the number of the valence $s$-quarks \cite{WangZG-Review,
 Wang-tetra-NPB-HCss,WangZG-IJMPA-2021,
 WangZG-XQ-mole-EPJA,Pcs4338-mole-XWWang}.

In Refs.\cite{Pcs4338-mole-XWWang,Pcs4459-mole-WangZG-SR,Pc4312-mole-penta-WXW-SCPMA,
Pc4312-mole-penta-WXW-IJMPA}, we resorted to  the scenario of hadronic molecules, see the assignments in Eq.\eqref{Assign-MB},
 constructed  the color singlet-singlet type  currents with definite isospins firstly, and studied the hidden-charm pentaquark (molecular) states with the  strangeness $S=0$ and $-1$ in the framework of the QCD sum rules  systematically with the help of the modified energy scale formula, and observed  that   except for the $P_c(4337)$ other pentaquark candidates could find their reasonable  rooms.
It is the unique/distinguished  feature of our works, which differs from the scheme of choosing the input parameters in other works \cite{Penta-mole-QCDSR-ChenHX-EPJC-2016,Penta-mole-QCDSR-Azizi-PRD-2017,
Penta-mole-QCDSR-Azizi-EPJC-2018,Penta-mole-QCDSR-ZhangJR-EPJC-2019,
Penta-mole-QCDSR-ChenHX-PRD-2019,Penta-mole-QCDSR-ChenHX-EPJC-2021,
Penta-mole-QCDSR-Ozdem-EPJC-2021,Penta-mole-QCDSR-LiuYL-EPJC-2021,
Penta-QCDSR-Azizi-EPJC-2022}, for detailed discussions about  this subject, see Ref.\cite{WangZG-Review}.

 In Refs.\cite{Wang1508-EPJC,WangHuang-EPJC-1508-12,WangZG-EPJC-1509-12,
 WangZG-NPB-1512-32,WangZhang-APPB}, we adopted  the pentaquark scenario in the diquark model  and carried out the operator product expansion up to the vacuum condensates of dimension 10 to  investigate  the diquark-diquark-antiquark type hidden-charm pentaquark states with the spin-parity  $J^P={\frac{1}{2}}^\pm$, ${\frac{3}{2}}^\pm$, ${\frac{5}{2}}^\pm$  and strangeness   $S=0,\,-1,\,-2,\,-3$ in the framework of  the QCD sum rules  systematically. Those works should be updated as the higher dimensional vacuum condensates with  $D>10$ are of great importance to obtain  flat Borel  platforms.

After the discovery of the $P_c(4312)$ \cite{LHCb-Pc4312},  we updated some old calculations by computing the  vacuum condensates up to dimension $13$  consistently,  and  restudied the ground state mass spectrum of the diquark-diquark-antiquark type $uudc\bar{c}$ pentaquark states with  the isospins $I=\frac{1}{2}$ and $\frac{3}{2}$ (but could not exhaust all the stable quark configurations), and identified  the $P_c(4312)$, $P_c(4380)$, $P_c(4440)$ and $P_c(4457)$ as compact pentaquark states  reasonably \cite{WZG-penta-IJMPA}. In Ref.\cite{WangZG-Pc12-Jpsip}, we  exhausted   the lowest  diquark-diquark-antiquark type $uudc\bar{c}$ pentaquark configurations with the isospin  $I=\frac{1}{2}$ (postponed the corresponding stable configurations with the isospin  $I=\frac{3}{2}$ to another work), and studied the mass spectrum  in the framework of the QCD sum rules in a comprehensive way, and revisited  the  identifications  of the $P_{c}$ states with the quantum numbers $IJ^P=\frac{1}{2}{\frac{1}{2}}^-$, $\frac{1}{2}{\frac{3}{2}}^-$ or $\frac{1}{2}{\frac{5}{2}}^-$.  We emphasize that we obtain more flat Borel windows in Refs.\cite{WZG-penta-IJMPA,WangZG-Pc12-Jpsip} as the higher dimensional vacuum condensates are of great importance  to obtain  flat  Borel windows, which live up to our naive expectations.

 After the discovery of the $P_{cs}(4459)$, we explored the possibility that  the $P_{c}(4312)$ and $P_{cs}(4459)$ are  isospin cousins with $I=\frac{1}{2}$ and $0$, respectively \cite{WangZG-Pcs4459-333}. In Ref.\cite{WangZG-Pc12-JpsiLambda},  we exhausted   the lowest  diquark-diquark-antiquark type $udsc\bar{c}$ configurations with the quantum numbers $IJ^P=0{\frac{1}{2}}^-$, $0{\frac{3}{2}}^-$ and $0{\frac{5}{2}}^-$ comprehensively,  and make possible identifications  of the $P_{cs}(4338)$ and $P_{cs}(4459)$ as compact pentaquark states  in a consistent way.

If the two-body strong decays $P_c \to J/\psi p$ and $P_{cs}\to J/\psi\Lambda$  conserve isospins,  the $P_c$ states have the isospin $I=\frac{1}{2}$ while the $P_{cs}$ states have the isospin $I=0$. However, we cannot exclude the possibility that the $P_c(4440)$ and $P_c(4457)$ have the isospin $I=\frac{3}{2}$, thus the strong decays $P_c(4440/4457) \to J/\psi p$ violate isospins, which can account for the narrow widths of the $P_c(4440/4457)$ naturally \cite{WangZG-Review,WangZG-Pc12-Jpsip}.  In Ref.\cite{WangZG-Pc12-JpsiXi}, we studied the lowest (the most stable)   diquark-diquark-antiquark type $qssc\bar{c}$ configurations with the quantum numbers $IJ^P=\frac{1}{2}{\frac{1}{2}}^-$, $\frac{1}{2}{\frac{3}{2}}^-$ and $\frac{1}{2}{\frac{5}{2}}^-$ comprehensively in the framework of  the QCD sum rules, which are expected to decay to the final states  $ J/\psi \Xi$  through the Okubo-Zweig-Iizuka super-allowed fall-apart mechanism.

 In this work,  we would like to study the lowest (the most stable)  diquark-diquark-antiquark type $uusc\bar{c}$ configurations with the quantum numbers $IJ^P=1{\frac{1}{2}}^-$, $1{\frac{3}{2}}^-$ and $1{\frac{5}{2}}^-$ comprehensively in the framework of  the QCD sum rules, as the observations  of their isospin cousins are of crucial importance. The light baryons $\Lambda$ and $\Sigma$ are isospin cousins with $I=0$ and $1$, respectively. The masses $M_{\Lambda^0}=1116\,\rm{MeV}$ and $M_{\Sigma^0}=1193\,\rm{MeV}$ from the Particle Data Group respectively \cite{PDG}, the isospin breaking effects are about $(M_{\Sigma^0}-M_{\Lambda^0})/(M_{\Sigma^0}+M_{\Lambda^0})=3\%$, the isospin is a good quantum number.

After the observations of the hidden-charm pentaquark candidates with the strangeness $S=0$ and $-1$ in the $J/\psi p$ and $J/\psi\Lambda $ invariant mass distributions respectively, it is natural and interesting to
search for the hidden-charm pentaquark candidates with the strangeness $S=-1$ in the $J/\psi\Sigma$ and $J/\psi \Sigma^*$ invariant mass distributions,  $S=-2$ in the $J/\psi\Xi$, $J/\psi\Xi^\prime$ and $J/\psi \Xi^*$ invariant mass distributions or $S=-3$ in the $J/\psi\Omega$ invariant mass distributions \cite{WangZG-Review,WangZG-EPJC-1509-12,WangZG-NPB-1512-32,di-di-anti-penta-1,di-di-anti-penta-2,Pc4312-penta-1}.

 The article is arranged as this way:  we obtain the QCD sum rules for the masses and pole residues of  the hidden-charm pentaquark states with strangeness in Sect.2;  in Sect.3, we present the numerical results and discussions; and Sect.4 is reserved for our
conclusion.

\section{QCD sum rules for  the  $uusc\bar{c}$ pentaquark states}
Routinely, we write down  the two-point correlation functions $\Pi(p)$, $\Pi_{\mu\nu}(p)$ and $\Pi_{\mu\nu\alpha\beta}(p)$ in the QCD sum rules,
\begin{eqnarray}\label{CF-Pi-Pi-Pi}
\Pi(p)&=&i\int d^4x e^{ip \cdot x} \langle0|T\left\{J(x)\bar{J}(0)\right\}|0\rangle \, ,\nonumber\\
\Pi_{\mu\nu}(p)&=&i\int d^4x e^{ip \cdot x} \langle0|T\left\{J_{\mu}(x)\bar{J}_{\nu}(0)\right\}|0\rangle \, ,\nonumber\\
\Pi_{\mu\nu\alpha\beta}(p)&=&i\int d^4x e^{ip \cdot x} \langle0|T\left\{J_{\mu\nu}(x)\bar{J}_{\alpha\beta}(0)\right\}|0\rangle \, ,
\end{eqnarray}
where the five-quark currents,
 \begin{eqnarray}
 J(x)&=&J^1(x)\, , \, J^2(x)\, , \, J^3(x)\, , \, J^4(x)\, , \, J^5(x)\, , \, J^6(x)\, , \nonumber\\
 J_\mu(x)&=&J_\mu^1(x)\, , \, J_\mu^2(x)\, , \, J_\mu^3(x)\, , \, J_\mu^4(x)\, , \, J_\mu^5(x)\, , \, J_\mu^6(x)\, , \, J_\mu^7(x)\, ,  \nonumber\\
 J_{\mu\nu}(x)&=&J_{\mu\nu}^1(x)\, , \, J_{\mu\nu}^2(x)\, ,\, J_{\mu\nu}^3(x)\, ,\, J_{\mu\nu}^4(x)\, ,\, J_{\mu\nu}^5(x)\, ,
 \end{eqnarray}
 with
\begin{eqnarray}\label{Current-12}
J^1(x)&=&\varepsilon^{ila} \varepsilon^{ijk}\varepsilon^{lmn}  s^T_j(x) C\gamma_5 u_k(x)\,u^T_m(x) C\gamma_5 c_n(x)\, C\bar{c}^{T}_{a}(x) \, , \nonumber \\
J^2(x)&=&\varepsilon^{ila} \varepsilon^{ijk}\varepsilon^{lmn}  s^T_j(x) C\gamma_5 u_k(x)\,u^T_m(x) C\gamma_\mu c_n(x)\,\gamma_5 \gamma^\mu C\bar{c}^{T}_{a}(x) \, ,\nonumber \\
J^{3}(x)&=&\frac{\varepsilon^{ila} \varepsilon^{ijk}\varepsilon^{lmn}}{\sqrt{2}} \left[ u^T_j(x) C\gamma_\mu u_k(x)s^T_m(x) C\gamma^\mu c_n(x)-u^T_j(x) C\gamma_\mu s_k(x)u^T_m(x) C\gamma^\mu c_n(x) \right]  C\bar{c}^{T}_{a}(x) \, , \nonumber\\
J^{4}(x)&=&\frac{\varepsilon^{ila} \varepsilon^{ijk}\varepsilon^{lmn}}{\sqrt{2}} \left[ u^T_j(x) C\gamma_\mu u_k(x) s^T_m(x) C\gamma_5 c_n(x)-u^T_j(x) C\gamma_\mu s_k(x) u^T_m(x) C\gamma_5 c_n(x)\right] \gamma_5 \gamma^\mu  C\bar{c}^{T}_{a}(x) \, ,  \nonumber\\
 J^{5}(x)&=&\varepsilon^{ila} \varepsilon^{ijk}\varepsilon^{lmn}  u^T_j(x) C\gamma_\mu u_k(x)\, s^T_m(x) C\gamma^\mu c_n(x) C\bar{c}^{T}_{a}(x) \, , \nonumber\\
J^{6}(x)&=&\varepsilon^{ila} \varepsilon^{ijk}\varepsilon^{lmn}  u^T_j(x) C\gamma_\mu u_k(x) \,s^T_m(x) C\gamma_5 c_n(x)  \gamma_5 \gamma^\mu C\bar{c}^{T}_{a}(x) \, ,
\end{eqnarray}
 for the isospin-spin $(I,J)=(1,\frac{1}{2})$,
\begin{eqnarray}\label{Current-32}
 J^1_{\mu}(x)&=&\varepsilon^{ila} \varepsilon^{ijk}\varepsilon^{lmn}
 s^T_j(x) C\gamma_5u_k(x)\,u^T_m(x) C\gamma_\mu c_n(x)\, C\bar{c}^{T}_{a}(x) \, , \nonumber \\
J^{2}_{\mu}(x)&=&\frac{\varepsilon^{ila} \varepsilon^{ijk}\varepsilon^{lmn}}{\sqrt{2}} \left[ u^T_j(x) C\gamma_\mu u_k(x) s^T_m(x) C\gamma_5 c_n(x) -u^T_j(x) C\gamma_\mu s_k(x) u^T_m(x) C\gamma_5 c_n(x)\right]   C\bar{c}^{T}_{a}(x) \, , \nonumber \\
 J^{3}_{\mu}(x)&=&\frac{\varepsilon^{ila} \varepsilon^{ijk}\varepsilon^{lmn}}{\sqrt{2}} \left[ u^T_j(x) C\gamma_\mu u_k(x)s^T_m(x) C\gamma_\alpha c_n(x)-u^T_j(x) C\gamma_\mu s_k(x)u^T_m(x) C\gamma_\alpha c_n(x) \right] \gamma_5\gamma^\alpha C\bar{c}^{T}_{a}(x) \, , \nonumber\\
J^{4}_{\mu}(x)&=&\frac{\varepsilon^{ila} \varepsilon^{ijk}\varepsilon^{lmn}}{\sqrt{2}} \left[ u^T_j(x) C\gamma_\alpha u_k(x)s^T_m(x) C\gamma_\mu c_n(x)-u^T_j(x) C\gamma_\alpha s_k(x)u^T_m(x) C\gamma_\mu c_n(x) \right] \gamma_5\gamma^\alpha C\bar{c}^{T}_{a}(x) \, ,\nonumber\\
 J^{5}_{\mu}(x)&=&\varepsilon^{ila} \varepsilon^{ijk}\varepsilon^{lmn} \ u^T_j(x) C\gamma_\mu u_k(x)\, s^T_m(x) C\gamma_5 c_n(x)    C\bar{c}^{T}_{a}(x) \, , \nonumber \\
J^{6}_{\mu}(x)&=&\varepsilon^{ila} \varepsilon^{ijk}\varepsilon^{lmn}  u^T_j(x) C\gamma_\mu u_k(x)\, s^T_m(x) C\gamma_\alpha c_n(x)\gamma_5\gamma^\alpha C\bar{c}^{T}_{a}(x) \, , \nonumber\\
J^{7}_{\mu}(x)&=&\varepsilon^{ila} \varepsilon^{ijk}\varepsilon^{lmn}  u^T_j(x) C\gamma_\alpha u_k(x)\, s^T_m(x) C\gamma_\mu c_n(x) \gamma_5\gamma^\alpha C\bar{c}^{T}_{a}(x) \, ,
\end{eqnarray}
 for the isospin-spin $(I,J)=(1,\frac{3}{2})$,
\begin{eqnarray} \label{Current-52}
J^1_{\mu\nu}(x)&=&\frac{\varepsilon^{ila} \varepsilon^{ijk}\varepsilon^{lmn} }{\sqrt{2}} s^T_j(x) C\gamma_5  u_k(x) \, u^T_m(x) C\gamma_\mu c_n(x)\, \gamma_5\gamma_{\nu}C\bar{c}^{T}_{a}(x)+(\mu \leftrightarrow \nu) \, ,\nonumber\\
J^2_{\mu\nu}(x)&=&\frac{\varepsilon^{ila} \varepsilon^{ijk}\varepsilon^{lmn} }{2}\, u^T_j(x) C\gamma_\mu u_k(x)\, s^T_m(x) C\gamma_5 c_n(x)\, \gamma_5\gamma_{\nu}C\bar{c}^{T}_{a}(x)+(\mu \leftrightarrow \nu) \, ,\nonumber\\
&&-\frac{\varepsilon^{ila} \varepsilon^{ijk}\varepsilon^{lmn} }{2}\, u^T_j(x) C\gamma_\mu s_k(x)\, u^T_m(x) C\gamma_5 c_n(x)\, \gamma_5\gamma_{\nu}C\bar{c}^{T}_{a}(x)+(\mu \leftrightarrow \nu) \, ,\nonumber\\
J^3_{\mu\nu}(x)&=&\frac{\varepsilon^{ila} \varepsilon^{ijk}\varepsilon^{lmn}}{2} u^T_j(x) C\gamma_\mu u_k(x)\, s^T_m(x) C\gamma_\nu c_n(x)  C\bar{c}^{T}_{a}(x)+(\mu \leftrightarrow \nu)\, ,\nonumber\\
&&-\frac{\varepsilon^{ila} \varepsilon^{ijk}\varepsilon^{lmn}}{2} u^T_j(x) C\gamma_\mu s_k(x)\, u^T_m(x) C\gamma_\nu c_n(x)  C\bar{c}^{T}_{a}(x)+(\mu \leftrightarrow \nu)\, ,\nonumber\\
J^4_{\mu\nu}(x)&=&\frac{\varepsilon^{ila} \varepsilon^{ijk}\varepsilon^{lmn} }{\sqrt{2}}\, s^T_j(x) C\gamma_\mu u_k(x)\, u^T_m(x) C\gamma_5 c_n(x)\, \gamma_5\gamma_{\nu}C\bar{c}^{T}_{a}(x)+(\mu \leftrightarrow \nu) \, ,\nonumber\\
J^5_{\mu\nu}(x)&=&\frac{\varepsilon^{ila} \varepsilon^{ijk}\varepsilon^{lmn}}{\sqrt{2}} u^T_j(x) C\gamma_\mu u_k(x)\, s^T_m(x) C\gamma_\nu c_n(x)  C\bar{c}^{T}_{a}(x)+(\mu \leftrightarrow \nu)\, ,
\end{eqnarray}
for the isospin-spin $(I,J)=(1,\frac{5}{2})$,
 the $i$, $j$, $k$, $l$, $m$, $n$ and $a$ are color indices, the $C$ is the charge conjugation matrix.

For the five-quark systems with the symbolic valence quarks $qqqQ\bar{Q}$, where $q=u$, $d$ or $s$, $Q=b$ or $c$, the valence quarks $qqq$ have the light-flavor $SU(3)$ symmetry while the valence quarks $Q\bar{Q}$ have the heavy quark symmetry. Thus,
\begin{eqnarray}\label{two-octet}
{\mathbf{3}}\otimes {\mathbf{3}}\otimes {\mathbf{3}} &\to & \left({\bar{\mathbf{3}}} \oplus {\mathbf{6}} \right) \otimes {\mathbf{3}}\, , \nonumber\\
&\to& \left({\mathbf{1}} \oplus {\mathbf{8}}_1\right) \oplus \left({\mathbf{8}}_2\oplus {\mathbf{10}}\right)\, ,
\end{eqnarray}
there exist  two octets, which could mix with each other. Experimentally,  the light baryons only have one octet, however, the $\Lambda(1405)$ maybe have two structures, i.e. $\Lambda(1380)$ and $\Lambda(1405)$, from the chiral unitary approach  \cite{Messiner-two-octet} (also see the subsection "Pole Structure of the $\Lambda(1405)$ Region" in  The Review of Particle Physics \cite{PDG}). we cannot  exclude another octet. The currents $J^{j=1-4}(x)$, $J^{j=1-4}_\mu(x)$ and $J^{j=1-3}_{\mu\nu}(x)$ belong to the octet ${\mathbf{8}}_1$ \cite{WangZG-Pc12-Jpsip,WangZG-Pc12-JpsiLambda,WangZG-Pc12-JpsiXi}, while the currents $J^{j=5-6}(x)$, $J^{j=5-7}_\mu(x)$ and $J^{j=4-5}_{\mu\nu}(x)$ belong to the octet ${\mathbf{8}}_2$ \cite{WangZG-Pc12-JpsiXi}. With the simple replacements,
\begin{eqnarray}
uus \to dds\, , \, \frac{ud+du}{\sqrt{2}}s \, ,
\end{eqnarray}
we obtain the corresponding currents with the symbolic valence quarks   $dds c\bar{c}$ and $\frac{uds c\bar{c}+dus c\bar{c}}{\sqrt{2}}$, and they couple potentially to the hidden-charm-singly-strange pentaquark states with degenerated masses.  For the $qqqc\bar{c}$ currents of the diquark-diquark-antiquark type in the $\mathbf{10}$ representation, we can consult Refs.\cite{WangZG-Review,WangZG-EPJC-1509-12,WangZG-NPB-1512-32,WZG-penta-IJMPA}.

Now we  analyze the quark structures of the currents $J(x)$, $J_\mu(x)$ and $J_{\mu\nu}(x)$ explicitly in the color space. The light (L) diquark operators $\varepsilon^{ijk}u^T_jC\gamma_{5}s_k$, $\varepsilon^{ijk}u^T_jC\gamma_{\mu}s_k$ and  $\varepsilon^{ijk}u^T_jC\gamma_{\mu}u_k$ have the spins $S_L=0$, $1$ and $1$, respectively, the heavy (H) diquark operators $\varepsilon^{lmn}u^T_mC\gamma_5c_n$, $\varepsilon^{lmn}s^T_mC\gamma_5c_n$, $\varepsilon^{lmn}u^T_mC\gamma_{\mu}c_n$ and   $\varepsilon^{lmn}s^T_mC\gamma_{\mu}c_n$ have the spins $S_H=0$, $0$, $1$ and $1$, respectively.   A light and  a heavy diquark form a tetraquark in the color $\mathbf{3}$ with  angular momentum $\vec{J}_{LH}=\vec{S}_L+\vec{S}_H$ having  values $J_{LH}=0$, $1$ or $2$.
The anti-charm quark operators $C\bar{c}_a^T$ and $\gamma_5\gamma_{\mu}C\bar{c}_a^T$ have the spin-parity $J^P={\frac{1}{2}}^-$ and ${\frac{3}{2}}^-$, respectively. The tensor $\varepsilon^{ila}$ in Eqs.\eqref{Current-12}-\eqref{Current-52} indicates the currents are of the diquark-diquark-antiquark type. As a result,  the total angular momentums (in other words, spins)  $\vec{J}=\vec{J}_{LH}+\vec{J}_{\bar{c}}$ have the values $J=\frac{1}{2}$, $\frac{3}{2}$ or $\frac{5}{2}$, the quark structures and corresponding spin-party  are shown clearly in Table \ref{current-pentaQ}.

\begin{table}
\begin{center}
\begin{tabular}{|c|c|c|c|c|c|c|c|c|}\hline\hline
$[qq][qc]\bar{c}$ ($S_L$, $S_H$, $J_{LH}$, $J$)  & $J^{P}$             & Currents              \\ \hline

$[su][uc]\bar{c}$ ($0$, $0$, $0$, $\frac{1}{2}$) &${\frac{1}{2}}^{-}$  &$J^1(x)$     \\

$[su][uc]\bar{c}$ ($0$, $1$, $1$, $\frac{1}{2}$) &${\frac{1}{2}}^{-}$  &$J^2(x)$    \\

$[uu][sc]\bar{c}-[us][uc]\bar{c}$ ($1$, $1$, $0$, $\frac{1}{2}$) &${\frac{1}{2}}^{-}$  &$J^3(x)$        \\

$[uu][sc]\bar{c}-[us][uc]\bar{c}$ ($1$, $0$, $1$, $\frac{1}{2}$) &${\frac{1}{2}}^{-}$  &$J^4(x)$             \\

$[uu][sc]\bar{c}$ ($1$, $1$, $0$, $\frac{1}{2}$) &${\frac{1}{2}}^{-}$  &$J^5(x)$        \\

$[uu][sc]\bar{c}$ ($1$, $0$, $1$, $\frac{1}{2}$) &${\frac{1}{2}}^{-}$  &$J^6(x)$             \\ \hline

$[su][uc]\bar{c}$ ($0$, $1$, $1$, $\frac{3}{2}$) &${\frac{3}{2}}^{-}$ &$J^1_\mu(x)$  \\

$[uu][sc]\bar{c}-[us][uc]\bar{c}$ ($1$, $0$, $1$, $\frac{3}{2}$) &${\frac{3}{2}}^{-}$ &$J^2_\mu(x)$          \\

$[uu][sc]\bar{c}-[us][uc]\bar{c}$ ($1$, $1$, $2$, $\frac{3}{2}$)${}_3$ &${\frac{3}{2}}^{-}$  &$J^3_\mu(x)$   \\

$[uu][sc]\bar{c}-[us][uc]\bar{c}$ ($1$, $1$, $2$, $\frac{3}{2}$)${}_4$ &${\frac{3}{2}}^{-}$  &$J^4_\mu(x)$   \\

$[uu][sc]\bar{c}$ ($1$, $0$, $1$, $\frac{3}{2}$) &${\frac{3}{2}}^{-}$ &$J^5_\mu(x)$          \\

$[uu][sc]\bar{c}$ ($1$, $1$, $2$, $\frac{3}{2}$)${}_6$ &${\frac{3}{2}}^{-}$  &$J^6_\mu(x)$   \\

$[uu][sc]\bar{c}$ ($1$, $1$, $2$, $\frac{3}{2}$)${}_7$ &${\frac{3}{2}}^{-}$  &$J^7_\mu(x)$   \\ \hline

$[su][uc]\bar{c}$ ($0$, $1$, $1$, $\frac{5}{2}$) &${\frac{5}{2}}^{-}$  &$J^1_{\mu\nu}(x)$     \\

$[uu][sc]\bar{c}-[us][uc]\bar{c}$ ($1$, $0$, $1$, $\frac{5}{2}$) &${\frac{5}{2}}^{-}$  &$J^2_{\mu\nu}(x)$    \\

$[uu][sc]\bar{c}-[us][uc]\bar{c}$ ($1$, $1$, $2$, $\frac{5}{2}$) &${\frac{5}{2}}^{-}$  &$J^3_{\mu\nu}(x)$   \\

$[su][uc]\bar{c}$ ($1$, $0$, $1$, $\frac{5}{2}$) &${\frac{5}{2}}^{-}$  &$J^4_{\mu\nu}(x)$    \\

$[uu][sc]\bar{c}$ ($1$, $1$, $2$, $\frac{5}{2}$) &${\frac{5}{2}}^{-}$  &$J^5_{\mu\nu}(x)$   \\
\hline\hline
\end{tabular}
\end{center}
\caption{ The valence quarks and spin-parity of the  currents, where the subscripts $3$, $4$, $6$ and $7$ denote the  superscripts of the corresponding currents.  }\label{current-pentaQ}
\end{table}

The coupling between the currents and baryons are complex due to the fraction spins.  The currents $J(x)$, $J_\mu(x)$ and $J_{\mu\nu}(x)$ have negative parity and  couple potentially to the hidden-charm-singly-strange  pentaquark states (P) with negative  and positive parities simultaneously as multiplying  $i\gamma_5$ to the interpolating  currents could change their parities \cite{WangZG-Review,Wang1508-EPJC},
\begin{eqnarray}\label{Coupling12}
\langle 0| J (0)|P_{\frac{1}{2}}^{-}(p)\rangle &=&\lambda^{-}_{\frac{1}{2}} U^{-}(p,s) \, , \nonumber \\
\langle 0| J (0)|P_{\frac{1}{2}}^{+}(p)\rangle &=&\lambda^{+}_{\frac{1}{2}} i\gamma_5 U^{+}(p,s) \, ,
\end{eqnarray}
\begin{eqnarray}
\langle 0| J_{\mu} (0)|P_{\frac{3}{2}}^{-}(p)\rangle &=&\lambda^{-}_{\frac{3}{2}} U^{-}_\mu(p,s) \, ,  \nonumber \\
\langle 0| J_{\mu} (0)|P_{\frac{3}{2}}^{+}(p)\rangle &=&\lambda^{+}_{\frac{3}{2}}i\gamma_5 U^{+}_\mu(p,s) \, ,  \nonumber \\
\langle 0| J_{\mu} (0)|P_{\frac{1}{2}}^{+}(p)\rangle &=&f^{+}_{\frac{1}{2}}p_\mu U^{+}(p,s) \, , \nonumber \\
\langle 0| J_{\mu} (0)|P_{\frac{1}{2}}^{-}(p)\rangle &=&f^{-}_{\frac{1}{2}}p_\mu i\gamma_5 U^{-}(p,s) \, ,
\end{eqnarray}
\begin{eqnarray}\label{Coupling52}
\langle 0| J_{\mu\nu} (0)|P_{\frac{5}{2}}^{-}(p)\rangle &=&\sqrt{2}\lambda^{-}_{\frac{5}{2}} U^{-}_{\mu\nu}(p,s) \, ,\nonumber\\
\langle 0| J_{\mu\nu} (0)|P_{\frac{5}{2}}^{+}(p)\rangle &=&\sqrt{2}\lambda^{+}_{\frac{5}{2}}i\gamma_5 U^{+}_{\mu\nu}(p,s) \, ,\nonumber\\
\langle 0| J_{\mu\nu} (0)|P_{\frac{3}{2}}^{+}(p)\rangle &=&f^{+}_{\frac{3}{2}} \left[p_\mu U^{+}_{\nu}(p,s)+p_\nu U^{+}_{\mu}(p,s)\right] \, , \nonumber\\
\langle 0| J_{\mu\nu} (0)|P_{\frac{3}{2}}^{-}(p)\rangle &=&f^{-}_{\frac{3}{2}}i\gamma_5 \left[p_\mu U^{-}_{\nu}(p,s)+p_\nu U^{-}_{\mu}(p,s)\right] \, , \nonumber\\
\langle 0| J_{\mu\nu} (0)|P_{\frac{1}{2}}^{-}(p)\rangle &=&g^{-}_{\frac{1}{2}}p_\mu p_\nu U^{-}(p,s) \, , \nonumber\\
\langle 0| J_{\mu\nu} (0)|P_{\frac{1}{2}}^{+}(p)\rangle &=&g^{+}_{\frac{1}{2}}p_\mu p_\nu i\gamma_5 U^{+}(p,s) \, ,
\end{eqnarray}
where  the superscripts $\pm$  represent   the  pentaquark parities, the subscripts $\frac{1}{2}$, $\frac{3}{2}$ and $\frac{5}{2}$  represent  the pentaquark spins,    the $\lambda$, $f$ and $g$ are the pentaquark pole residues, in other words, the current-hadron couplings. The currents $J(x)$ couple potentially to the pentaquark states with the spin-parity $J^P={\frac{1}{2}}^{\pm}$, the currents $J_\mu(x)$ couple potentially to the pentaquark states with the spin-parity $J^P={\frac{3}{2}}^{\pm}$ and ${\frac{1}{2}}^{\pm}$,
the currents $J_{\mu\nu}(x)$ couple potentially to the pentaquark states with the spin-parity $J^P={\frac{5}{2}}^{\pm}$, ${\frac{3}{2}}^{\pm}$ and ${\frac{1}{2}}^{\pm}$.
The $U^\pm(p,s)$,  $U^{\pm}_\mu(p,s)$ and $U^{\pm}_{\mu\nu}(p,s)$ are Dirac and Rarita-Schwinger spinors respectively \cite{WangZG-Review,Wang1508-EPJC}.
Although there exists  arbitrariness in the Lagrangians of the baryons (pentaquarks) with higher spins, the equation of motions known as Rarita-Schwinger framework could remove the redundancies efficiently to construct concise projectors $P^{\pm}_{\mu\nu}=\sum_s U^{\pm}_\mu(p,s) \bar{U}^{\pm}_\mu(p,s)$, $P^{\pm}_{\mu\nu\alpha\beta}=\sum_s U^{\pm}_{\mu\nu}(p,s) \bar{U}^{\pm}_{\alpha\beta}(p,s)$, etc \cite{RS-EPJA,HuangShiZhong}.

On the other hand, if we perform Firez transformation both in the color and Dirac spinor spaces, we can transform  the currents $J(x)$, $J_\mu(x)$ and $J_{\mu\nu}(x)$ into a series of color singlet-singlet type currents. For example,
\begin{eqnarray}\label{Firez-Trans}
8J^1&=& 2iS_{su} c\, \bar{c} i\gamma_5 u
-2S_{su} \gamma_5 c\, \bar{c} u +2S_{su} \gamma_{\alpha} c\, \bar{c} \gamma^{\alpha}\gamma_5 u
+2S_{su} \gamma_{\alpha}\gamma_5 c\,\bar{c} \gamma^{\alpha} u\nonumber\\
&&+S_{su} \sigma_{\alpha\beta}\gamma_5 c\, \bar{c} \sigma^{\alpha\beta} u -2iS_{su} u\, \bar{c} i\gamma_5 c+2S_{su} \gamma_5 u\, \bar{c} c
-2S_{su} \gamma_{\alpha} u\, \bar{c} \gamma^{\alpha}\gamma_5 c\nonumber\\
&&-2S_{su} \gamma_{\alpha}\gamma_5 u\, \bar{c} \gamma^{\alpha} c
-S_{su} \sigma_{\alpha\beta}\gamma_5 u\, \bar{c} \sigma^{\alpha\beta} c\, ,
\end{eqnarray}
where $S_{su}\Gamma c=\varepsilon^{ijk}s^T_i C\gamma_5 u_j \Gamma c_k$,
$S_{su}\Gamma u=\varepsilon^{ijk}s^T_i C\gamma_5 u_j \Gamma u_k$, and the $\Gamma$ denotes some Dirac $\gamma$-matrixes. The components, in other words, the color singlet-singlet type local currents, couple potentially to the compact pentaquark states with two valence color-neutral clusters, which have the same quantum numbers as the physical  mesons and baryons except for the masses. Or the diquark-diquark-antiquark type pentaquark states have many compact (not loose) color singlet-singlet type Fock components and embody the net effects.  The decays to the physical meson-baryon pairs could take place with the Okubo-Zweig-Iizuka super-allowed fall-apart mechanism, if they are kinematically allowed. Although the color singlet-singlet type pentaquark states are usually called molecular states in the QCD sum rules, they are compact objects  according to local currents, it is a unique feature of the QCD sum rules.

At the hadron  side, we insert  a complete set  of intermediate
hidden-charm-singly-strange  pentaquark states with the same quantum numbers as the currents  $J(x)$, $i\gamma_5 J(x)$, $J_{\mu}(x)$, $i\gamma_5 J_{\mu}(x)$, $J_{\mu\nu}(x)$  and $i\gamma_5 J_{\mu\nu}(x)$ into the correlation functions
$\Pi(p)$, $\Pi_{\mu\nu}(p)$ and $\Pi_{\mu\nu\alpha\beta}(p)$ to obtain the hadronic representation
\cite{SVZ79-1,SVZ79-2,PRT85},  isolate the  lowest  states, and obtain the results\footnote{Here we present  an example to illustrate the technical details,
\begin{eqnarray}
\Pi(p)&=&i\int d^4x e^{ip\cdot x} \int \frac{d^3\vec{k}}{(2\pi)^3 2\omega_k}  \left[e^{+i\vec{k} \cdot \vec{x}}  e^{-i\omega_k t}\theta(t)+ e^{-i\vec{k} \cdot \vec{x}}e^{+i\omega_k t}\theta(-t)\right]\, {\rm sum}+\cdots\, , \nonumber\\
&=&i\int d^4x e^{ip\cdot x} e^{i\vec{k} \cdot \vec{x}}\int \frac{d^3\vec{k}}{(2\pi)^3 2\omega_k}  \left[  e^{-i\omega_k t}\theta(t)+ e^{+i\omega_k t}\theta(-t)\right]\, {\rm sum}+ \cdots\, , \nonumber\\
&=&\lim_{\varepsilon \to 0}i\int d^4x e^{ip\cdot x} e^{i\vec{k} \cdot \vec{x}}\int \frac{d^3\vec{k}}{(2\pi)^3 2\omega_k}  \frac{-2\omega_k}{2\pi i}\int_{-\infty}^{+\infty}\frac{d\omega}{\omega^2-\omega_k^2+i\varepsilon} e^{-i\omega t} \, {\rm sum}+\cdots\, , \nonumber\\
&=&-\lim_{\varepsilon \to 0}\int \frac{d^4k}{(2\pi)^4 }   (2\pi)^4 \,\delta^4(p-k)\, {\lambda^{-}_{\frac{1}{2}}}^2 \frac{\!\not\!{k}+M_{-}}{k^2-M_{-}^2+i\varepsilon} +  \cdots\, , \nonumber\\
&=& {\lambda^{-}_{\frac{1}{2}}}^2 \frac{\!\not\!{p}+M_{-}}{M_{-}^2-p^2} +  \cdots\, ,
\end{eqnarray}
where
${\rm sum}=\sum  \langle0|J(0)|P^-_{\frac{1}{2}}(k)\rangle  \langle P^-_{\frac{1}{2}}(k)|\bar{J}(0) |0\rangle$.}
\begin{eqnarray}\label{CF-Hadron-12}
\Pi(p) & = & {\lambda^{-}_{\frac{1}{2}}}^2  {\!\not\!{p}+ M_{-} \over M_{-}^{2}-p^{2}  }+  {\lambda^{+}_{\frac{1}{2}}}^2  {\!\not\!{p}- M_{+} \over M_{+}^{2}-p^{2}  } +\cdots  \, ,\nonumber\\
&=&\Pi_{\frac{1}{2}}^1(p^2)\!\not\!{p}+\Pi_{\frac{1}{2}}^0(p^2)\, ,
 \end{eqnarray}
\begin{eqnarray}\label{CF-Hadron-32}
 \Pi_{\mu\nu}(p) & = & {\lambda^{-}_{\frac{3}{2}}}^2  {\!\not\!{p}+ M_{-} \over M_{-}^{2}-p^{2}  } \left(- g_{\mu\nu}+\frac{\gamma_\mu\gamma_\nu}{3}+\frac{2p_\mu p_\nu}{3p^2}-\frac{p_\mu\gamma_\nu-p_\nu \gamma_\mu}{3\sqrt{p^2}}
\right)\nonumber\\
&&+  {\lambda^{+}_{\frac{3}{2}}}^2  {\!\not\!{p}- M_{+} \over M_{+}^{2}-p^{2}  } \left(- g_{\mu\nu}+\frac{\gamma_\mu\gamma_\nu}{3}+\frac{2p_\mu p_\nu}{3p^2}-\frac{p_\mu\gamma_\nu-p_\nu \gamma_\mu}{3\sqrt{p^2}}
\right)   \nonumber \\
& &+ {f^{+}_{\frac{1}{2}}}^2  {\!\not\!{p}+ M_{+} \over M_{+}^{2}-p^{2}  } p_\mu p_\nu+  {f^{-}_{\frac{1}{2}}}^2  {\!\not\!{p}- M_{-} \over M_{-}^{2}-p^{2}  } p_\mu p_\nu  +\cdots  \, , \nonumber\\
&=&\left[\Pi_{\frac{3}{2}}^1(p^2)\!\not\!{p}+\Pi_{\frac{3}{2}}^0(p^2)\right]\left(- g_{\mu\nu}\right)+\cdots\, ,
\end{eqnarray}
\begin{eqnarray}\label{CF-Hadron-52}
\Pi_{\mu\nu\alpha\beta}(p) & = &2{\lambda^{-}_{\frac{5}{2}}}^2  {\!\not\!{p}+ M_{-} \over M_{-}^{2}-p^{2}  } \left[\frac{ \widetilde{g}_{\mu\alpha}\widetilde{g}_{\nu\beta}+\widetilde{g}_{\mu\beta}\widetilde{g}_{\nu\alpha}}{2}-\frac{\widetilde{g}_{\mu\nu}\widetilde{g}_{\alpha\beta}}{5}-\frac{1}{10}\left( \gamma_{\mu}\gamma_{\alpha}+\frac{\gamma_{\mu}p_{\alpha}-\gamma_{\alpha}p_{\mu}}{\sqrt{p^2}}-\frac{p_{\mu}p_{\alpha}}{p^2}\right)\widetilde{g}_{\nu\beta}\right.\nonumber\\
&&\left.-\frac{1}{10}\left( \gamma_{\nu}\gamma_{\alpha}+\frac{\gamma_{\nu}p_{\alpha}-\gamma_{\alpha}p_{\nu}}{\sqrt{p^2}}-\frac{p_{\nu}p_{\alpha}}{p^2}\right)\widetilde{g}_{\mu\beta}
+\cdots\right]\nonumber\\
&&+  2 {\lambda^{+}_{\frac{5}{2}}}^2  {\!\not\!{p}- M_{+} \over M_{+}^{2}-p^{2}  } \left[\frac{ \widetilde{g}_{\mu\alpha}\widetilde{g}_{\nu\beta}+\widetilde{g}_{\mu\beta}\widetilde{g}_{\nu\alpha}}{2}
-\frac{\widetilde{g}_{\mu\nu}\widetilde{g}_{\alpha\beta}}{5}-\frac{1}{10}\left( \gamma_{\mu}\gamma_{\alpha}+\frac{\gamma_{\mu}p_{\alpha}-\gamma_{\alpha}p_{\mu}}{\sqrt{p^2}}-\frac{p_{\mu}p_{\alpha}}{p^2}\right)\widetilde{g}_{\nu\beta}\right.\nonumber\\
&&\left.
-\frac{1}{10}\left( \gamma_{\nu}\gamma_{\alpha}+\frac{\gamma_{\nu}p_{\alpha}-\gamma_{\alpha}p_{\nu}}{\sqrt{p^2}}-\frac{p_{\nu}p_{\alpha}}{p^2}\right)\widetilde{g}_{\mu\beta}
 +\cdots\right]   \nonumber\\
 && +{f^{+}_{\frac{3}{2}}}^2  {\!\not\!{p}+ M_{+} \over M_{+}^{2}-p^{2}  } \left[ p_\mu p_\alpha \left(- g_{\nu\beta}+\frac{\gamma_\nu\gamma_\beta}{3}+\frac{2p_\nu p_\beta}{3p^2}-\frac{p_\nu\gamma_\beta-p_\beta \gamma_\nu}{3\sqrt{p^2}}
\right)+\cdots \right]\nonumber\\
&&+  {f^{-}_{\frac{3}{2}}}^2  {\!\not\!{p}- M_{-} \over M_{-}^{2}-p^{2}  } \left[ p_\mu p_\alpha \left(- g_{\nu\beta}+\frac{\gamma_\nu\gamma_\beta}{3}+\frac{2p_\nu p_\beta}{3p^2}-\frac{p_\nu\gamma_\beta-p_\beta \gamma_\nu}{3\sqrt{p^2}}
\right)+\cdots \right]   \nonumber \\
& &+ {g^{-}_{\frac{1}{2}}}^2  {\!\not\!{p}+ M_{-} \over M_{-}^{2}-p^{2}  } p_\mu p_\nu p_\alpha p_\beta+  {g^{+}_{\frac{1}{2}}}^2  {\!\not\!{p}- M_{+} \over M_{+}^{2}-p^{2}  } p_\mu p_\nu p_\alpha p_\beta  +\cdots \, , \nonumber\\
& = & \left[\Pi_{\frac{5}{2}}^1(p^2)\!\not\!{p}+\Pi_{\frac{5}{2}}^0(p^2)\right]\left( g_{\mu\alpha}g_{\nu\beta}+g_{\mu\beta}g_{\nu\alpha}\right)  +\cdots \, ,
 \end{eqnarray}
where $\widetilde{g}_{\mu\nu}=g_{\mu\nu}-\frac{p_{\mu}p_{\nu}}{p^2}$, the $M_{\mp}$ are the  masses of the hidden-charm pentaquark states with the negative and positive parities, respectively, and permutations of the Lorentz indexes $\mu$, $\nu$, $\alpha$ and $\beta$ are implied. We adopt  the components $\Pi_{\frac{1}{2}}^1(p^2)$, $\Pi_{\frac{1}{2}}^0(p^2)$, $\Pi_{\frac{3}{2}}^1(p^2)$, $\Pi_{\frac{3}{2}}^0(p^2)$, $\Pi_{\frac{5}{2}}^1(p^2)$ and  $\Pi_{\frac{5}{2}}^0(p^2)$ to avoid  contaminations.

We should bear in mind that the quantum field theory does not prohibit  the non-vanishing couplings of the currents $J(x)$, $J_{\mu}(x)$ and $J_{\mu\nu}(x)$ to the meson-baryon scattering states $\bar{D}\Xi_c$, $\bar{D}^*\Xi_c$, $\bar{D}_s\Sigma_c$, $\bar{D}_s^*\Sigma_c$,
$J/\psi \Sigma$, $\eta_c \Sigma$, $\cdots$ if they have the isospin, spin and parity, according to the Firez transformation (see Eq.\eqref{Firez-Trans}) and related discussions.
We choose the local five-quark currents, see Eqs.\eqref{Current-12}-\eqref{Current-52}, while the conventional mesons (M) and baryons (B) are spatial extended objects and have finite mean spatial sizes $\sqrt{\langle r^2\rangle} \neq 0$, and the meson-baryon pairs (or loose molecular states from the quark models, heavy meson/baryon chiral theory) have large  sizes $r=\sqrt{\langle r^2\rangle_M}+\sqrt{\langle r^2\rangle_B}\geq 1\,\rm{fm}$,   the couplings are expected to be very small indeed \cite{WangZG-Review}. Direct investigations indicate that the two-hadron (meson-meson, meson-baryon, baryon-antibaryon, etc) scattering states alone  cannot saturate the QCD sum rules, their net effects could be absorbed into the pole residues with a simple redefinition safely. All in all,  the two-particle scattering states are of minor
importance in the QCD sum rules if we choose local five-quark currents \cite{WangZG-Review,WangZG-landau-PRD,WangZG-IJMPA-Two-part}.

Then we routinely obtain the  spectral densities  through  dispersion relation,
\begin{eqnarray}
\frac{{\rm Im}\Pi^1_j(s)}{\pi}&=& \lambda_{-}^2 \delta\left(s-M_{-}^2\right)+\lambda_{+}^2 \delta\left(s-M_{+}^2\right) =\, \rho^1_{H}(s) \, , \\
\frac{{\rm Im}\Pi^0_j(s)}{\pi}&=&M_{-}\lambda_{-}^2 \delta\left(s-M_{-}^2\right)-M_{+}\lambda_{+}^2 \delta\left(s-M_{+}^2\right)
=\rho^0_{H}(s) \, ,
\end{eqnarray}
where the subscripts $j=\frac{1}{2}$, $\frac{3}{2}$, $\frac{5}{2}$, the subscript $H$  stands for  the hadron side, the $\lambda_{\mp}$ are  the pole residues of the pentaquark states with the negative and positive parities respectively, and we have neglected  the spin indexes $\frac{1}{2}$, $\frac{3}{2}$, $\frac{5}{2}$ for simplicity.
We resort to the  weight functions $\sqrt{s}\exp\left(-\frac{s}{T^2}\right)$ and $\exp\left(-\frac{s}{T^2}\right)$ to obtain the QCD sum rules,
\begin{eqnarray}
\int_{4m_c^2}^{s_0}ds \left[\sqrt{s}\,\rho^1_{H}(s)+\rho^0_{H}(s)\right]\exp\left( -\frac{s}{T^2}\right)
&=&2M_{-}\lambda_{-}^2\exp\left( -\frac{M_{-}^2}{T^2}\right) \, ,
\end{eqnarray}
\begin{eqnarray}
\int_{4m_c^2}^{s^\prime_0}ds \left[\sqrt{s}\,\rho^1_{H}(s)-\rho^0_{H}(s)\right]\exp\left( -\frac{s}{T^2}\right)
&=&2M_{+}\lambda_{+}^2\exp\left( -\frac{M_{+}^2}{T^2}\right) \, ,
\end{eqnarray}
where the $s_0$ and $s_0^\prime$ are the continuum thresholds for the $uusc\bar{c}$ pentaquark states with the negative and
positive parities, respectively. The pentaquark states with negative and positive parities receive no contaminations from each other \cite{WangZG-Review,Wang1508-EPJC,WangHuang-EPJC-1508-12,WangZG-EPJC-1509-12,WangZG-NPB-1512-32,
WangZhang-APPB}.

At the QCD side,  we carry out  the operator product expansion with the help of the full $u$, $d$, $s$ and $c$ quark propagators routinely \cite{WangZG-Review,Wang-tetra-NPB-HCss,WangZG-IJMPA-2021,
WZG-penta-IJMPA,WangZG-Pcs4459-333,Wang-tetra-PRD-HC,
WZG-tetraquark-Mc}, and compute  the quark-gluon operators up to dimension $13$ and order $\mathcal{O}( \alpha_s^{k})$ with $k\leq 1$ in a consistent way to acquire  the vacuum condensates, and   take account of the terms  $\propto m_s$ according to  the light-flavor   $SU(3)$ mass-breaking effects \cite{WangZG-Review}.
The higher dimensional  vacuum condensates $\langle\bar{q}q\rangle\langle\bar{q}g_s\sigma Gq\rangle\langle\frac{\alpha_sGG}{\pi}\rangle$,  $\langle\bar{q}g_s\sigma Gq\rangle^2\langle\frac{\alpha_sGG}{\pi}\rangle$, $\langle\bar{q}g_s\sigma Gq\rangle^3$ and $\langle\bar{q}q\rangle^2\langle\bar{q}g_s\sigma Gq\rangle\langle\frac{\alpha_sGG}{\pi}\rangle$ have the dimensions $n=12$, $14$, $15$ and $15$, respectively, but they come from the quark-gluon operators of the orders $\mathcal{O}( \alpha_s^{k})$ with $k=\frac{3}{2}$, $2$, $\frac{3}{2}$ and $\frac{3}{2}$, respectively. At the typical energy scale $\mu=2.5\,\rm{GeV}$, $\alpha_s\approx 0.27$, their contributions are greatly depressed and can be neglected safely \cite{WangZG-Review}.

Again, we routinely obtain  the spectral densities through   dispersion relation,
\begin{eqnarray}\label{QCD-rho}
 \rho^1_{QCD}(s) &=&\frac{{\rm Im}\Pi^1_j(s)}{\pi}\, , \nonumber\\
\rho^0_{QCD}(s) &=&\frac{{\rm Im}\Pi^0_j(s)}{\pi}\, .
\end{eqnarray}

Now we  match the hadron side with the QCD side of the correlation functions, take the quark-hadron duality below the continuum thresholds, and  obtain  two  QCD sum rules:
\begin{eqnarray}\label{QCDSR}
2M_{-}\lambda_{-}^2\exp\left( -\frac{M_{-}^2}{T^2}\right)&=& \int_{4m_c^2}^{s_0}ds \,\left[\sqrt{s}\rho_{QCD}^1(s)+\rho_{QCD}^{0}(s)\right]\,\exp\left( -\frac{s}{T^2}\right)\,  ,
\end{eqnarray}
\begin{eqnarray}\label{QCDSR-Positive}
2M_{+}\lambda_{+}^2\exp\left( -\frac{M_{+}^2}{T^2}\right)&=& \int_{4m_c^2}^{s^\prime_0}ds \,\left[\sqrt{s}\rho_{QCD}^1(s)-\rho_{QCD}^{0}(s)\right]\,\exp\left( -\frac{s}{T^2}\right)\,  .
\end{eqnarray}

If we  retain the  pole residues $\lambda_{\mp}$, we can obtain two conventional  QCD sum rules with respect to the components $\Pi_{j}^{1}(p^2)$ and $\Pi_{j}^{0}(p^2)$, respectively \cite{PRT85,QiaoCF-review},
\begin{eqnarray}\label{Traditional-QCDSR-1}
\lambda_{-}^2\exp\left( -\frac{M_{-}^2}{T^2}\right)+\lambda_{+}^2\exp\left( -\frac{M_{+}^2}{T^2}\right)&=&\int_{4m_c^2}^{s_0}ds \,\rho^1_{QCD}(s)\exp\left( -\frac{s}{T^2}\right) \, ,
\end{eqnarray}
\begin{eqnarray}\label{Traditional-QCDSR-0}
M_{-}\lambda_{-}^2\exp\left( -\frac{M_{-}^2}{T^2}\right)-M_{+}\lambda_{+}^2\exp\left( -\frac{M_{+}^2}{T^2}\right)&=&\int_{4m_c^2}^{s_0}ds \,\rho^0_{QCD}(s)\exp\left( -\frac{s}{T^2}\right) \, .
\end{eqnarray}
It is obvious that there are contributions come from the pentaquark states with the positive parity. If we neglect their contributions just by setting $\lambda_{+}=0$, their net effects would be embodied in the parameters $\lambda_{-}$ and $M_{-}$.
 In Refs.\cite{WangZG-Pc12-Jpsip,WangZG-Pc12-JpsiLambda,WangZG-Pc12-JpsiXi},  we suggest a parameter CTM to represent  contaminations from the hidden-charm (no-strange,  singly-strange, doubly-strange)  pentaquark states with the positive parity in the case of setting $\lambda_{+}=0$,
\begin{eqnarray}
{\rm CTM}&=&\frac{\int_{4m_c^2}^{s_0}ds \,\left[\sqrt{s}\rho_{QCD}^1(s)-\rho_{QCD}^{0}(s)\right]\,\exp\left( -\frac{s}{T^2}\right)}{\int_{4m_c^2}^{s_0}ds \,\left[\sqrt{s}\rho_{QCD}^1(s)+\rho_{QCD}^{0}(s)\right]\,\exp\left( -\frac{s}{T^2}\right)}\, ,
\end{eqnarray}
by setting $s^\prime_0=s_0$.
Direct calculations indicate that ${\rm CTM}\sim 0.10$ or $0.20$ in the Borel windows \cite{WangZG-Pc12-Jpsip,WangZG-Pc12-JpsiLambda,WangZG-Pc12-JpsiXi} (including the present calculations), the contaminations from the hidden-charm pentaquark states with positive parity are considerable.

In this work, we adopt the QCD sum rules for the hidden-charm-singly-strange pentaquark states $P_{cs}$ with negative parity, i.e. Eq.\eqref{QCDSR}, and differentiate   Eq.\eqref{QCDSR} in regard  to the variable  $\frac{1}{T^2}$, then eliminate the pole residues $\lambda_{-}$ through a fraction to obtain  the QCD sum rules for the  masses cleanly,
 \begin{eqnarray}
 M^2_{-} &=& \frac{-\int_{4m_c^2}^{s_0}ds \frac{d}{d(1/T^2)}\, \left[\sqrt{s}\rho_{QCD}^1(s)+\rho_{QCD}^{0}(s)\right]\,\exp\left( -\frac{s}{T^2}\right)}{\int_{4m_c^2}^{s_0}ds \, \left[\sqrt{s}\rho_{QCD}^1(s)+\rho_{QCD}^{0}(s)\right]\,\exp\left( -\frac{s}{T^2}\right)}\,  .
\end{eqnarray}

\section{Numerical results and discussions}
At the initial points, we take  the  vacuum condensates
$\langle\bar{q}q \rangle=-(0.24\pm 0.01\, \rm{GeV})^3$,  $\langle\bar{s}s \rangle=(0.8\pm0.1)\langle\bar{q}q \rangle$,
 $\langle\bar{q}g_s\sigma G q \rangle=m_0^2\langle \bar{q}q \rangle$, $\langle\bar{s}g_s\sigma G s \rangle=m_0^2\langle \bar{s}s \rangle$,
$m_0^2=(0.8 \pm 0.1)\,\rm{GeV}^2$, $\langle \frac{\alpha_s
GG}{\pi}\rangle=0.012\pm0.004\,\rm{GeV}^4$    at the energy scale  $\mu=1\, \rm{GeV}$
\cite{SVZ79-1,SVZ79-2,PRT85,ColangeloReview}, and  take the $\overline{MS}$  quark  masses $m_{c}(m_c)=(1.275\pm0.025)\,\rm{GeV}$
 and $m_s(\mu=2\,\rm{GeV})=(0.095\pm0.005)\,\rm{GeV}$
 from the Particle Data Group \cite{PDG}.
Furthermore,  we take into account  the energy-scale dependence  \cite{Narison-mix},
 \begin{eqnarray}
 \langle\bar{q}q \rangle(\mu)&=&\langle\bar{q}q\rangle({\rm 1 GeV})\left[\frac{\alpha_{s}({\rm 1 GeV})}{\alpha_{s}(\mu)}\right]^{\frac{12}{33-2n_f}}\, , \nonumber\\
 \langle\bar{s}s \rangle(\mu)&=&\langle\bar{s}s \rangle({\rm 1 GeV})\left[\frac{\alpha_{s}({\rm 1 GeV})}{\alpha_{s}(\mu)}\right]^{\frac{12}{33-2n_f}}\, , \nonumber\\
 \langle\bar{q}g_s \sigma Gq \rangle(\mu)&=&\langle\bar{q}g_s \sigma Gq \rangle({\rm 1 GeV})\left[\frac{\alpha_{s}({\rm 1 GeV})}{\alpha_{s}(\mu)}\right]^{\frac{2}{33-2n_f}}\, ,\nonumber\\
  \langle\bar{s}g_s \sigma Gs \rangle(\mu)&=&\langle\bar{s}g_s \sigma Gs \rangle({\rm 1 GeV})\left[\frac{\alpha_{s}({\rm 1 GeV})}{\alpha_{s}(\mu)}\right]^{\frac{2}{33-2n_f}}\, ,\nonumber\\
m_c(\mu)&=&m_c(m_c)\left[\frac{\alpha_{s}(\mu)}{\alpha_{s}(m_c)}\right]^{\frac{12}{33-2n_f}} \, ,\nonumber\\
m_s(\mu)&=&m_s({\rm 2GeV} )\left[\frac{\alpha_{s}(\mu)}{\alpha_{s}({\rm 2GeV})}\right]^{\frac{12}{33-2n_f}}\, ,\nonumber\\
\alpha_s(\mu)&=&\frac{1}{b_0t}\left[1-\frac{b_1}{b_0^2}\frac{\log t}{t} +\frac{b_1^2(\log^2{t}-\log{t}-1)+b_0b_2}{b_0^4t^2}\right]\, ,
\end{eqnarray}
  where $t=\log \frac{\mu^2}{\Lambda^2}$, $b_0=\frac{33-2n_f}{12\pi}$, $b_1=\frac{153-19n_f}{24\pi^2}$, $b_2=\frac{2857-\frac{5033}{9}n_f+\frac{325}{27}n_f^2}{128\pi^3}$,  $\Lambda_{QCD}=210\,\rm{MeV}$, $292\,\rm{MeV}$  and  $332\,\rm{MeV}$ for the quark  flavors  $n_f=5$, $4$ and $3$, respectively  \cite{PDG}.

As in previous works \cite{WangZG-Pc12-Jpsip,WangZG-Pc12-JpsiLambda,WangZG-Pc12-JpsiXi}, we study  the hidden-charm pentaquark states $qqqc\bar{c}$ (P) with $q=u$, $d$ or $s$,  and choose the quark flavor numbers $n_f=4$, then evolve all the input parameters to a typical energy scale $\mu$, which satisfies  the modified energy scale formula,
\begin{eqnarray}\label{formula}
\mu &=&\sqrt{M_{P}^2-(2{\mathbb{M}}_c)^2}-\kappa{\mathbb{M}}_s \, ,
 \end{eqnarray}
 with the effective quark masses ${\mathbb{M}}_c$ and ${\mathbb{M}}_s$, where the $\kappa$ is the number of the valence $s$-quarks.

Phenomenologically,  we  describe the hidden-heavy five-quark systems  $qqqQ\bar{Q}$  by a double-well potential. In the heavy quark limit, the $Q$ and $\bar{Q}$ serve as two static well potentials, the diquark $[qq]$ and quark $q$  lie in the two wells,  respectively. Then they form the diquark-diquark-antiquark type or color singlet-singlet type pentaquark states.
The five-quark systems  $qqqQ\bar{Q}$ are characterized by the effective heavy quark masses ${\mathbb{M}}_Q$  and
the virtuality $V=\sqrt{M^2_{P}-(2{\mathbb{M}}_Q)^2}$. It is natural to set the energy  scale  as $\mu=V$.
We usually set the small $u$ and $d$ quark masses to be zero, and take account of the light-flavor $SU(3)$
breaking effects by introducing an effective $s$-quark mass ${\mathbb{M}}_s$, thus we obtain the modified energy scale formula. It has been approved  that the (modified) energy scale formula can enhance the pole contributions  remarkably on the hadron side and improve the convergent behaviors of the operator product expansion remarkably on the QCD side.

We can rewrite the energy scale formula in Eq.\eqref{formula} in a Regge-like form,
\begin{eqnarray}\label{Modify-formula-Regge}
M^2_{X/Y/Z/P}&=&(\mu+\kappa {\mathbb{M}}_s)^2+{\rm Constants}\, .
\end{eqnarray}
where the Constants have the values $4{\mathbb{M}}_Q^2$.
  The ${\mathbb{M}}_Q$ and ${\mathbb{M}}_s$ are fitted by the QCD sum rules,  the  updated values are ${\mathbb{M}}_c=1.82\,\rm{GeV}$ and ${\mathbb{M}}_s=0.15\,\rm{GeV}$ respectively, and the modified energy scale formula works well for the hidden-charm/bottom (doubly-charm/bottom) tetraquark states, pentaquark states and molecular states \cite{WangZG-Review,Wang-tetra-formula,WangZG-mole-formula,
 Wang-tetra-NPB-HCss,WangZG-IJMPA-2021,WangZG-XQ-mole-EPJA,Pcs4338-mole-XWWang,
 WangZG-Pc12-Jpsip,WangZG-Pcs4459-333,
WangZG-Pc12-JpsiLambda,WangZG-Pc12-JpsiXi,
 Wang-tetra-PRD-HC,
 WZG-tetraquark-Mc}.

It is easy to obtain the relation among  uncertainties,
$M_P\delta M_P=(\mu+\kappa {\mathbb{M}}_s)(\delta\mu+\kappa \delta{\mathbb{M}}_s)+4{\mathbb{M}}_c \delta {\mathbb{M}}_c \approx \mu\delta\mu+4{\mathbb{M}}_c \delta {\mathbb{M}}_c$. A small uncertainty $\delta {\mathbb{M}}_c$ could be compensated by a uncertainty  $\delta\mu=-4\frac{{\mathbb{M}}_c}{\mu}\delta {\mathbb{M}}_c$ to warrant $\delta M_P=0$. In fact, there exists  a universal and  optimal ${\mathbb{M}}_c$ and we can set $\delta{\mathbb{M}}_c=0$, thus $M_P\delta M_P \approx \mu\delta\mu$. In calculations, we retain the precision up to $0.1\,\rm{GeV}$ for the energy scale $\mu$ via the round method, the uncertainty $\delta \mu<0.05\,\rm{GeV}$ lead to $\delta M_P=\frac{\mu}{M_P}\delta \mu <0.03\,\rm{GeV}$, and the tiny uncertainty  can be neglected.

In the QCD sum rules for  the  baryons  and  pentaquark (molecular) states contain at least one valence heavy quark,  we usually choose the continuum thresholds as $\sqrt{s_0}=M_{gr}+ (0.5-0.8)\,\rm{GeV}$  \cite{WangZG-Review,Pcs4338-mole-XWWang,Pc4312-mole-penta-WXW-SCPMA,
Pc4312-mole-penta-WXW-IJMPA,Wang1508-EPJC,WangHuang-EPJC-1508-12,
 WangZG-EPJC-1509-12,WangZG-NPB-1512-32,WZG-penta-IJMPA,
 WangZG-Pc12-Jpsip,WangZG-Pcs4459-333,WangZG-Pc12-JpsiLambda,WangZG-Pc12-JpsiXi,
 Wang-cc-baryon-penta},   where the subscript $gr$ stands for the ground states.
 In our previous works,  such a criterion worked  very well in interpreting  the $P_c(4312)$, $P_c(4337)$, $P_{cs}(4338)$, $P_{c}(4380)$, $P_c(4440)$, $P_c(4457)$ and $P_{cs}(4459)$ \cite{WangZG-Pc12-Jpsip,WangZG-Pc12-JpsiLambda,WangZG-Pc12-JpsiXi}.

After numerous  trial  and error, we acquire reasonable  Borel  windows and continuum threshold parameters in our unique scheme in treating the input parameters \cite{WangZG-Review}, which are shown explicitly in Table \ref{Borel}. From the table, we can see clearly   that the pole (or ground state) contributions are about $(40-60)\%$, the largest pole contributions up to now,  the pole dominance criterion is satisfied very well,
  routinely, the pole contributions are defined by,
\begin{eqnarray}
{\rm{pole}}&=&\frac{\int_{4m_{c}^{2}}^{s_{0}}ds\,\rho_{QCD}\left(s\right)\exp\left(-\frac{s}{T^{2}}\right)} {\int_{4m_{c}^{2}}^{\infty}ds\,\rho_{QCD}\left(s\right)\exp\left(-\frac{s}{T^{2}}\right)}\, ,
\end{eqnarray}
 with $\rho_{QCD}=\sqrt{s}\rho_{QCD}^1(s)+\rho_{QCD}^{0}(s)$.

\begin{figure}
\centering
\includegraphics[totalheight=6cm,width=7cm]{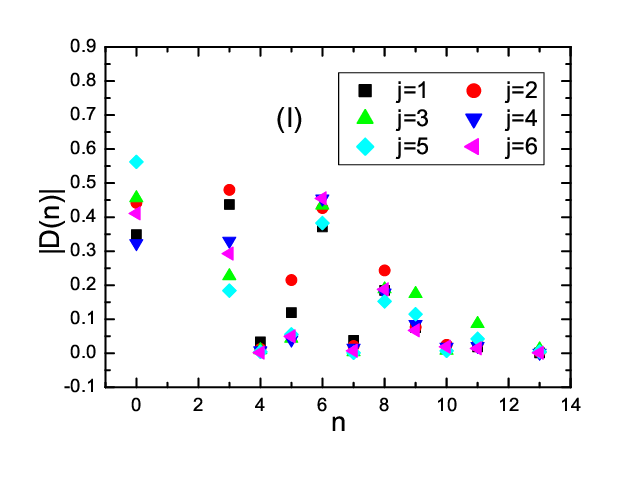}
\includegraphics[totalheight=6cm,width=7cm]{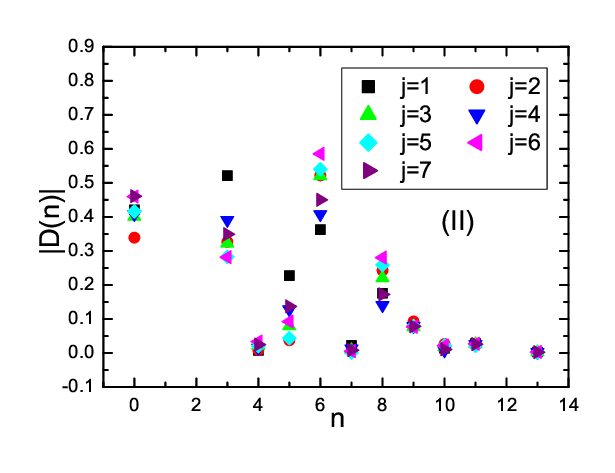}
\includegraphics[totalheight=6cm,width=7cm]{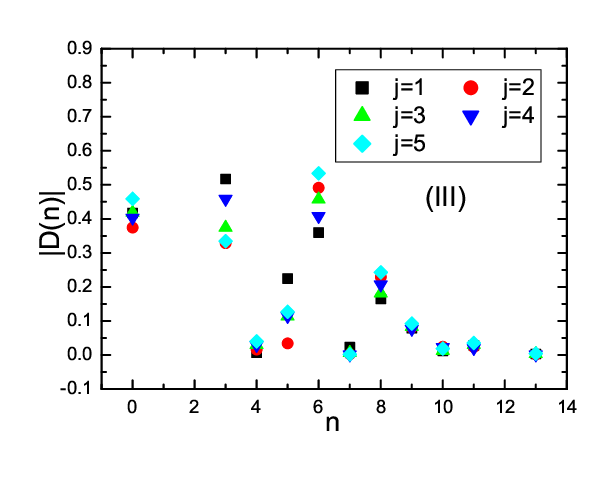}
 \caption{ The $|D(n)|$ with variations of the $n$ for the central values of the input parameters, where the (I), (II) and (III) represent   the spins  $J=\frac{1}{2}$, $\frac{3}{2}$ and $\frac{5}{2}$ of the currents respectively, the $j=1$, $2$, $3$, $4$, $5$, $6$ and $7$ represent the series numbers of the currents. }\label{OPE-fig}
\end{figure}

 In Fig.\ref{OPE-fig}, we plot the absolute values of the contributions of the vacuum condensates of dimension $n$ for  central values of all the input  parameters, again the $D(n)$ are routinely defined by,
   \begin{eqnarray}
D(n)&=&\frac{\int_{4m_{c}^{2}}^{s_{0}}ds\,\rho_{QCD,n}(s)\exp\left(-\frac{s}{T^{2}}\right)}
{\int_{4m_{c}^{2}}^{s_{0}}ds\,\rho_{QCD}\left(s\right)\exp\left(-\frac{s}{T^{2}}\right)}\, .
\end{eqnarray}
From the figure, we can see clearly  that the contributions  $D(4)$ and $D(7)$ are  neglectable, their tiny values even can be neglected  in determining the Borel windows,  while the $D(6)$ plays  a most  important role in most cases, and could serve as a milestone to judge the convergent behaviors. The contributions $|D(n)|$ of the vacuum condensates with $n\geq 6$ have the hierarchies,
\begin{eqnarray}
&&D(6)\gg |D(8)| \gg D(9) \gg D(10)\sim |D(11)| \gg D(13) \, ,
\end{eqnarray}
roughly,  the convergent behaviors of the operator product expansion  are very good. For readers  convenience, we write down the explicit forms of  those vacuum condensates with $n\geq 6$ in above equation,
\begin{eqnarray}
n=6&:&\langle \bar{q}q\rangle^2\, , \,\langle \bar{q}q\rangle\langle \bar{s}s\rangle\, ; \nonumber\\
n=8&:&\langle \bar{q}q\rangle\langle \bar{q}g_s\sigma Gq\rangle\, , \, \langle \bar{q}q\rangle\langle \bar{s}g_s\sigma Gs\rangle\, , \,\langle \bar{s}s\rangle\langle \bar{q}g_s\sigma Gq\rangle\,  ; \nonumber\\
n=9&:&\langle \bar{q}q\rangle^2\langle \bar{s}s\rangle\, ; \nonumber\\
n=10&:&\langle \bar{q}g_s\sigma Gq\rangle^2\, , \, \langle \bar{q}g_s\sigma Gq\rangle\langle \bar{s}g_s\sigma Gs\rangle\, , \,\langle \bar{q}q\rangle^2\langle \frac{\alpha_sGG}{\pi}\rangle\, , \,\langle \bar{q}q\rangle\langle \bar{s}s\rangle \langle \frac{\alpha_sGG}{\pi}\rangle\,  ; \nonumber\\
n=11&:&\langle \bar{q}q\rangle^2\langle \bar{s}g_s\sigma Gs\rangle\, , \, \langle \bar{q}q\rangle\langle \bar{s}s\rangle\langle \bar{q}g_s\sigma Gq\rangle\,  ; \nonumber\\
n=13&:&\langle\bar{s}s\rangle\langle \bar{q}g_s\sigma Gq\rangle^2\, , \, \langle\bar{q}q\rangle\langle \bar{q}g_s\sigma Gq\rangle\langle \bar{s}g_s\sigma Gs\rangle\, , \,\langle \bar{q}q\rangle^2\langle \bar{s}s\rangle\langle \frac{\alpha_sGG}{\pi}\rangle\,   .
\end{eqnarray}

\begin{table}
\begin{center}
\begin{tabular}{|c|c|c|c|c|c|c|c|}\hline\hline
                  &$T^2(\rm{GeV}^2)$     &$\sqrt{s_0}(\rm{GeV})$    &$\mu(\rm{GeV})$  &pole          &$D(13)$         \\ \hline

$J^1(x)$          &$3.4-3.8$             &$5.16\pm0.10$             &$2.5$            &$(41-61)\%$   &$\ll 1\%$      \\ \hline

$J^2(x)$          &$3.4-3.8$             &$5.25\pm0.10$             &$2.6$            &$(42-62)\%$   &$\ll 1\%$       \\ \hline

$J^3(x)$          &$3.0-3.4$             &$5.03\pm0.10$             &$2.2$            &$(41-63)\%$   &$< 2\%$      \\ \hline

$J^4(x)$          &$3.2-3.6$             &$5.03\pm0.10$             &$2.2$            &$(40-62)\%$   &$\ll1\%$     \\ \hline

$J^5(x)$          &$3.3-3.7$             &$5.20\pm0.10$             &$2.5$            &$(40-61)\%$   &$\ll 1\%$      \\ \hline

$J^6(x)$          &$3.4-3.8$             &$5.16\pm0.10$             &$2.4$            &$(40-60)\%$   &$\ll1\%$     \\ \hline

$J^1_\mu(x)$      &$3.5-3.9$             &$5.25\pm0.10$             &$2.6$            &$(41-60)\%$   &$\ll 1\%$     \\ \hline

$J^2_\mu(x)$      &$3.3-3.7$             &$5.05\pm0.10$             &$2.3$            &$(40-60)\%$   &$\ll1\%$     \\ \hline

$J^3_\mu(x)$      &$3.4-3.8$             &$5.16\pm0.10$             &$2.4$            &$(40-61)\%$   &$\ll1\%$     \\ \hline

$J^4_\mu(x)$      &$3.4-3.8$             &$5.16\pm0.10$             &$2.4$            &$(40-60)\%$   &$\ll1\%$     \\ \hline

$J^5_\mu(x)$      &$3.4-3.8$             &$5.17\pm0.10$             &$2.5$            &$(41-61)\%$   &$\ll1\%$     \\ \hline

$J^6_\mu(x)$      &$3.4-3.8$             &$5.20\pm0.10$             &$2.5$            &$(41-61)\%$   &$\ll1\%$     \\ \hline

$J^7_\mu(x)$      &$3.4-3.8$             &$5.20\pm0.10$             &$2.5$            &$(40-61)\%$   &$\ll1\%$     \\ \hline

$J^1_{\mu\nu}(x)$ &$3.5-3.9$             &$5.25\pm0.10$             &$2.6$            &$(41-61)\%$   &$\ll1\%$     \\ \hline

$J^2_{\mu\nu}(x)$ &$3.4-3.8$             &$5.14\pm0.10$             &$2.4$            &$(41-61)\%$   &$\ll1\%$     \\ \hline

$J^3_{\mu\nu}(x)$ &$3.5-3.9$             &$5.20\pm0.10$             &$2.5$            &$(40-60)\%$   &$\ll1\%$     \\ \hline

$J^4_{\mu\nu}(x)$ &$3.5-3.9$             &$5.25\pm0.10$             &$2.6$            &$(42-62)\%$   &$\ll1\%$     \\ \hline

$J^5_{\mu\nu}(x)$ &$3.4-3.8$             &$5.21\pm0.10$             &$2.5$            &$(42-62)\%$   &$\ll1\%$     \\ \hline

\hline
\end{tabular}
\end{center}
\caption{ The Borel  windows, continuum thresholds, optimal energy scales, pole contributions,   contributions of the vacuum condensates of dimension 13 for the hidden-charm-singly-strange pentaquark states with the isospin $I=1$. }\label{Borel}
\end{table}

\begin{table}
\begin{center}
\begin{tabular}{|c|c|c|c|c|c|c|c|c|}\hline\hline
$[qq][qc]\bar{c}$ ($S_L$, $S_H$, $J_{LH}$, $J$) &$M(\rm{GeV})$   &$\lambda(10^{-3}\rm{GeV}^6)$       \\ \hline

$[su][uc]\bar{c}$ ($0$, $0$, $0$, $\frac{1}{2}$)  &$4.48\pm0.10$ &$1.93\pm0.31$                  \\ \hline

$[su][uc]\bar{c}$ ($0$, $1$, $1$, $\frac{1}{2}$)  &$4.59\pm0.10$ &$3.72\pm0.58$                 \\ \hline

$[uu][sc]\bar{c}-[us][uc]\bar{c}$ ($1$, $1$, $0$, $\frac{1}{2}$)  &$4.35\pm0.11$ &$2.95\pm0.53$                \\ \hline

$[uu][sc]\bar{c}-[us][uc]\bar{c}$ ($1$, $0$, $0$, $\frac{1}{2}$)  &$4.35\pm0.11$ &$3.19\pm0.54$                \\ \hline

$[uu][sc]\bar{c}$ ($1$, $1$, $0$, $\frac{1}{2}$)  &$4.50\pm0.12$ &$4.66\pm0.82$                \\ \hline

$[uu][sc]\bar{c}$ ($1$, $0$, $0$, $\frac{1}{2}$)  &$4.47\pm0.11$ &$4.57\pm0.75$                \\ \hline

$[su][uc]\bar{c}$ ($0$, $1$, $1$, $\frac{3}{2}$)  &$4.58\pm0.10$ &$2.06\pm0.33$                \\ \hline

$[uu][sc]\bar{c}-[us][uc]\bar{c}$ ($1$, $0$, $1$, $\frac{3}{2}$)  &$4.38\pm0.10$ &$1.85\pm0.31$                \\ \hline

$[uu][sc]\bar{c}-[us][uc]\bar{c}$ ($1$, $1$, $2$, $\frac{3}{2}$)${}_3$ &$4.47\pm0.11$  &$3.74\pm0.60$   \\ \hline

$[uu][sc]\bar{c}-[us][uc]\bar{c}$ ($1$, $1$, $2$, $\frac{3}{2}$)${}_4$ &$4.47\pm0.11$   &$3.73\pm0.61$    \\ \hline

$[uu][sc]\bar{c}$ ($1$, $0$, $1$, $\frac{3}{2}$)  &$4.48\pm0.10$ &$2.56\pm0.41$                \\ \hline

$[uu][sc]\bar{c}$ ($1$, $1$, $2$, $\frac{3}{2}$)${}_6$ &$4.52\pm0.10$  &$4.58\pm0.73$   \\ \hline

$[uu][sc]\bar{c}$ ($1$, $1$, $2$, $\frac{3}{2}$)${}_7$ &$4.51\pm0.11$   &$4.56\pm0.74$    \\ \hline

$[su][uc]\bar{c}$ ($0$, $1$, $1$, $\frac{5}{2}$)  &$4.57\pm0.11$ &$2.05\pm0.32$                     \\ \hline

$[uu][sc]\bar{c}-[us][uc]\bar{c}$ ($1$, $0$, $1$, $\frac{5}{2}$)  &$4.45\pm0.10$ &$2.12\pm0.34$                     \\ \hline

$[uu][sc]\bar{c}-[us][uc]\bar{c}$ ($1$, $1$, $2$, $\frac{5}{2}$)  &$4.51\pm0.11$   &$2.20\pm0.35$                  \\ \hline

$[su][uc]\bar{c}$ ($1$, $0$, $1$, $\frac{5}{2}$)  &$4.56\pm0.11$ &$2.06\pm0.32$                     \\ \hline

$[uu][sc]\bar{c}$ ($1$, $1$, $2$, $\frac{5}{2}$)  &$4.52\pm0.11$   &$2.52\pm0.40$                  \\ \hline\hline
\end{tabular}
\end{center}
\caption{ The masses  and pole residues of the hidden-charm-singly-strange pentaquark states. }\label{mass-Pcs}
\end{table}

\begin{figure}
\centering
\includegraphics[totalheight=6cm,width=7cm]{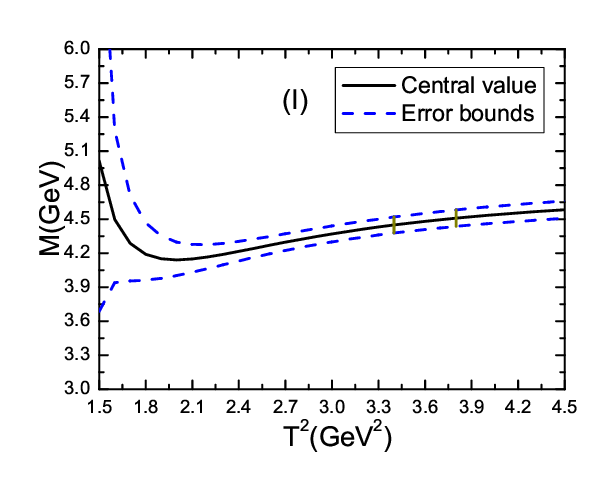}
\includegraphics[totalheight=6cm,width=7cm]{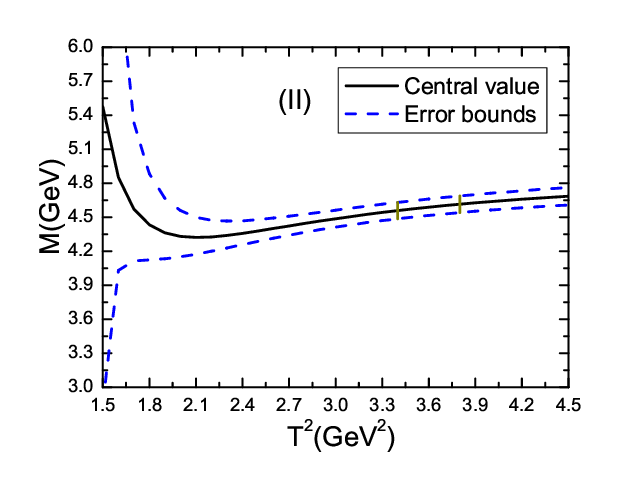}
\includegraphics[totalheight=6cm,width=7cm]{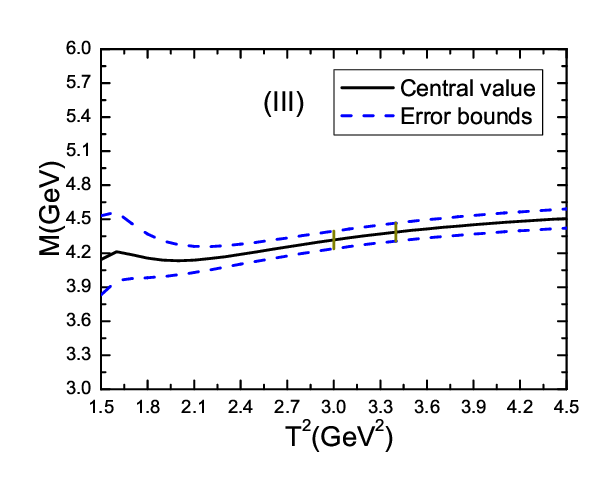}
\includegraphics[totalheight=6cm,width=7cm]{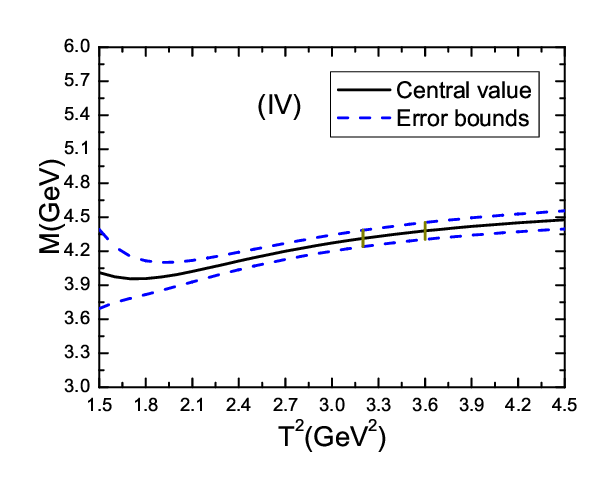}
\includegraphics[totalheight=6cm,width=7cm]{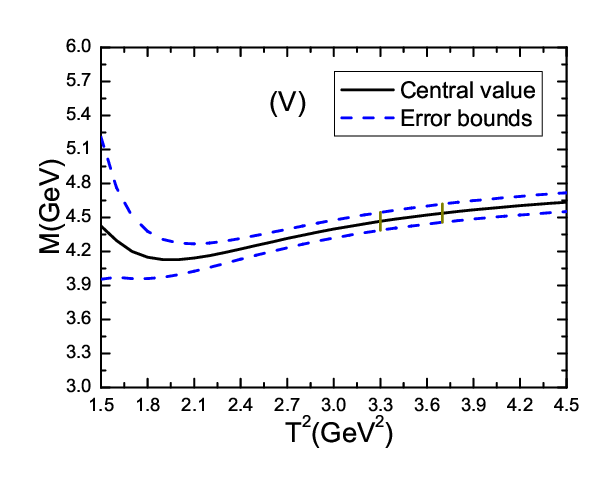}
\includegraphics[totalheight=6cm,width=7cm]{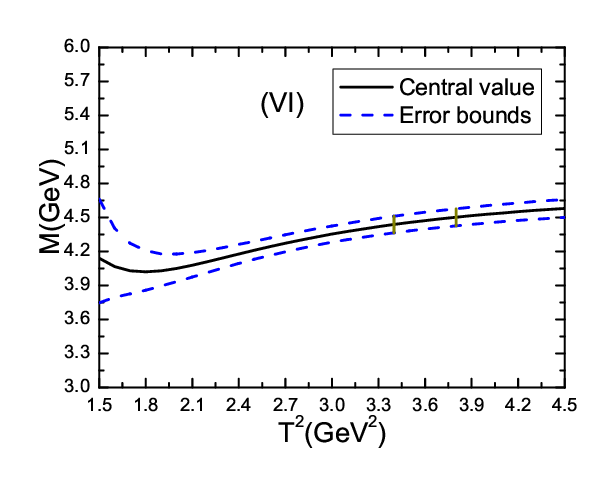}
  \caption{ The masses  with variations of the  Borel parameters $T^2$ for  the hidden-charm-singly-strange  pentaquark states, where the (I), (II), (III), (IV), (V)  and (VI)  represent the
   $[su][uc]\bar{c}$ ($0$, $0$, $0$, $\frac{1}{2}$),                    $[su][uc]\bar{c}$ ($0$, $1$, $1$, $\frac{1}{2}$), $[uu][sc]\bar{c}-[us][uc]\bar{c}$ ($1$, $1$, $0$, $\frac{1}{2}$),
$[uu][sc]\bar{c}-[us][uc]\bar{c}$ ($1$, $0$, $0$, $\frac{1}{2}$),
$[uu][sc]\bar{c}$ ($1$, $1$, $0$, $\frac{1}{2}$)   and
$[uu][sc]\bar{c}$ ($1$, $0$, $0$, $\frac{1}{2}$)  pentaquark states, respectively. }\label{mass-1-fig}
\end{figure}

\begin{figure}
\centering
\includegraphics[totalheight=6cm,width=7cm]{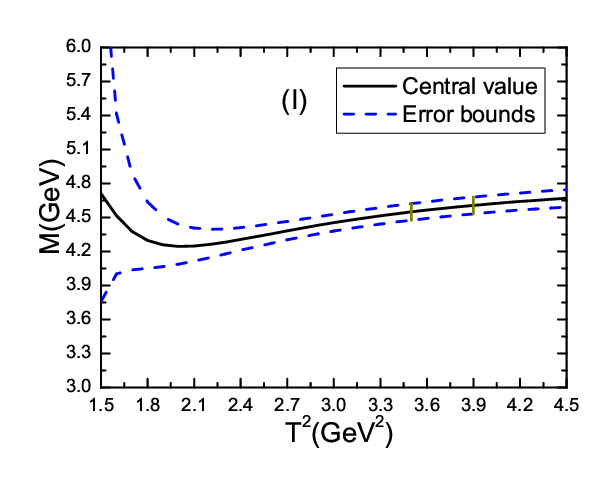}
\includegraphics[totalheight=6cm,width=7cm]{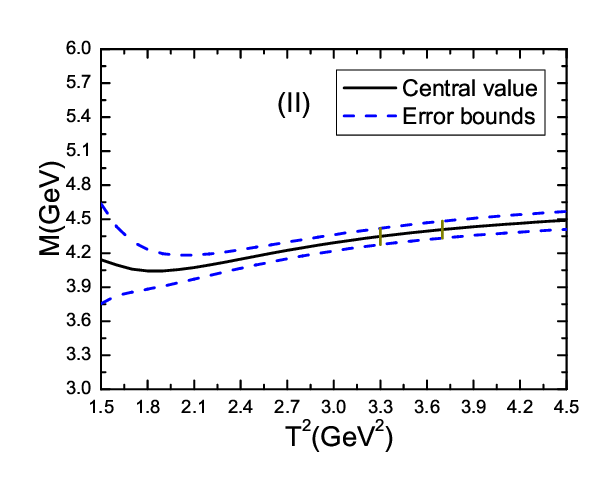}
\includegraphics[totalheight=6cm,width=7cm]{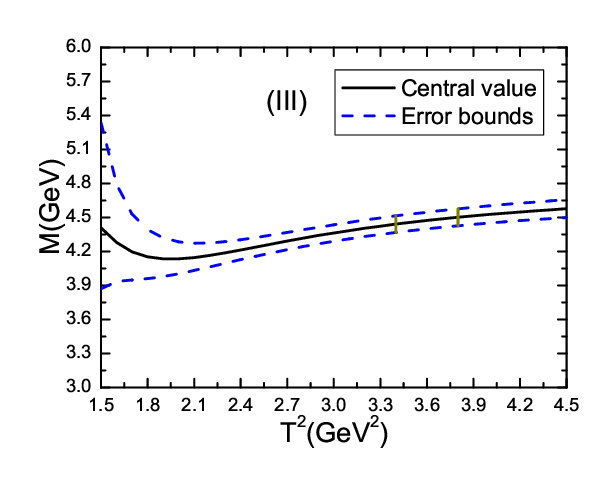}
\includegraphics[totalheight=6cm,width=7cm]{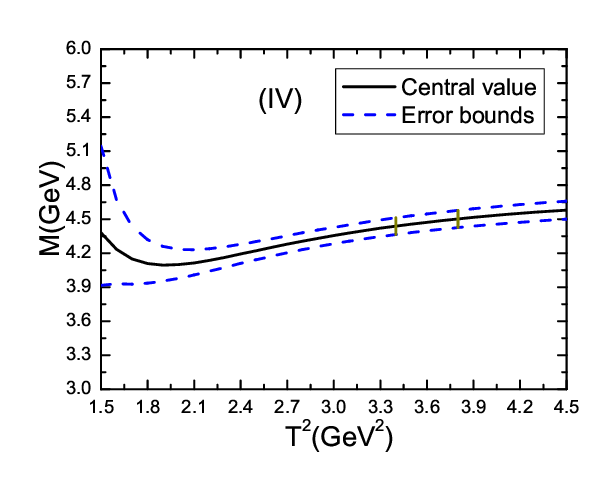}
\includegraphics[totalheight=6cm,width=7cm]{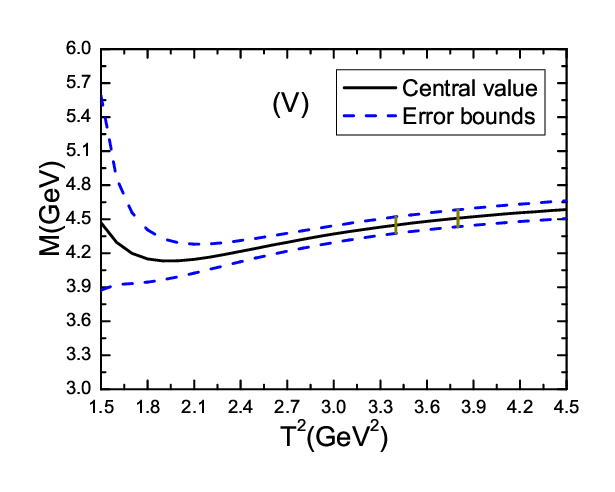}
\includegraphics[totalheight=6cm,width=7cm]{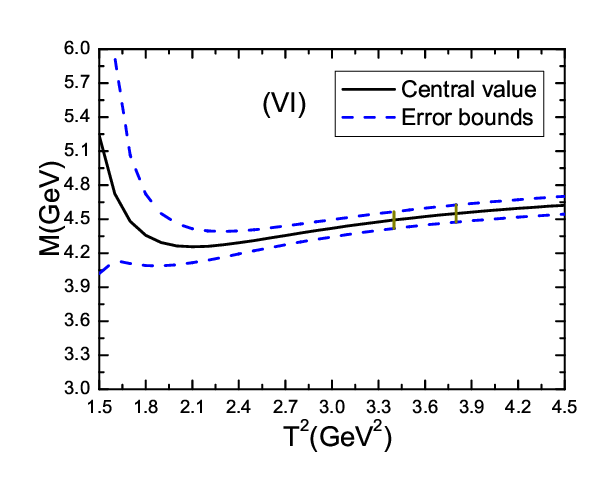}
  \caption{ The masses  with variations of the  Borel parameters $T^2$ for  the hidden-charm-singly-strange pentaquark states, where the (I), (II), (III), (IV), (V) and (VI) represent the $[su][uc]\bar{c}$ ($0$, $1$, $1$, $\frac{3}{2}$),
$[uu][sc]\bar{c}-[us][uc]\bar{c}$ ($1$, $0$, $1$, $\frac{3}{2}$),
$[uu][sc]\bar{c}-[us][uc]\bar{c}$ ($1$, $1$, $2$, $\frac{3}{2}$)${}_3$,
$[uu][sc]\bar{c}-[us][uc]\bar{c}$ ($1$, $1$, $2$, $\frac{3}{2}$)${}_4$, $[uu][sc]\bar{c}$ ($1$, $0$, $1$, $\frac{3}{2}$) and $[uu][sc]\bar{c}$ ($1$, $1$, $2$, $\frac{3}{2}$)${}_6$  pentaquark states, respectively.  }\label{mass-2-fig-1}
\end{figure}

\begin{figure}
\centering
\includegraphics[totalheight=6cm,width=7cm]{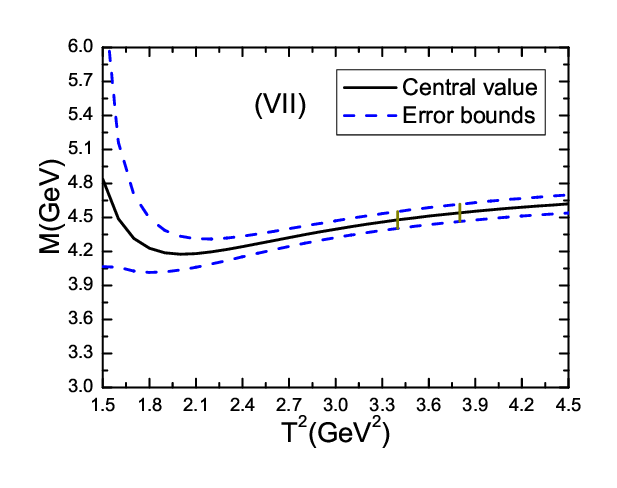}
  \caption{ The mass  with variations of the  Borel parameter $T^2$ for  the hidden-charm-singly-strange pentaquark state, where the (VII) represents the
$[uu][sc]\bar{c}$ ($1$, $1$, $2$, $\frac{3}{2}$)${}_7$ pentaquark state.  }\label{mass-2-fig-2}
\end{figure}

\begin{figure}
\centering
\includegraphics[totalheight=6cm,width=7cm]{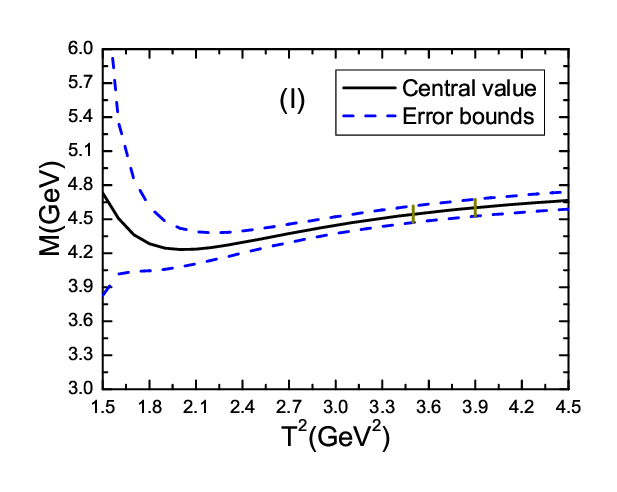}
\includegraphics[totalheight=6cm,width=7cm]{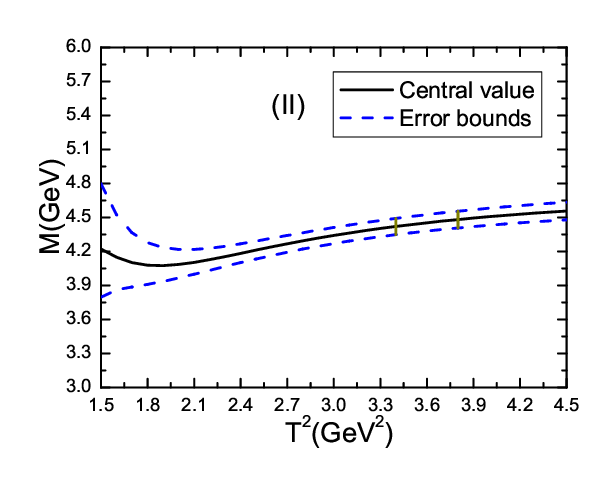}
\includegraphics[totalheight=6cm,width=7cm]{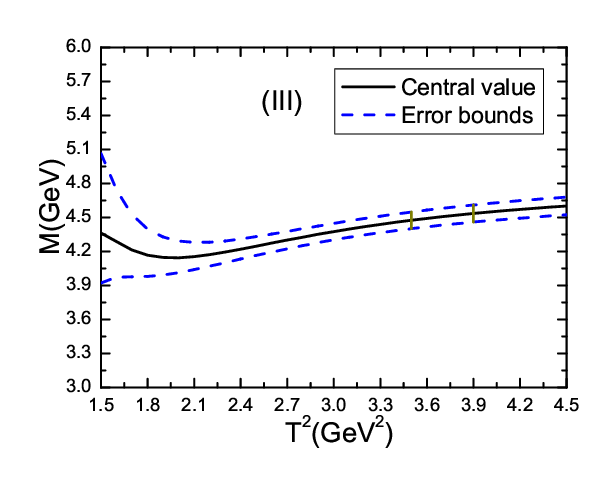}
\includegraphics[totalheight=6cm,width=7cm]{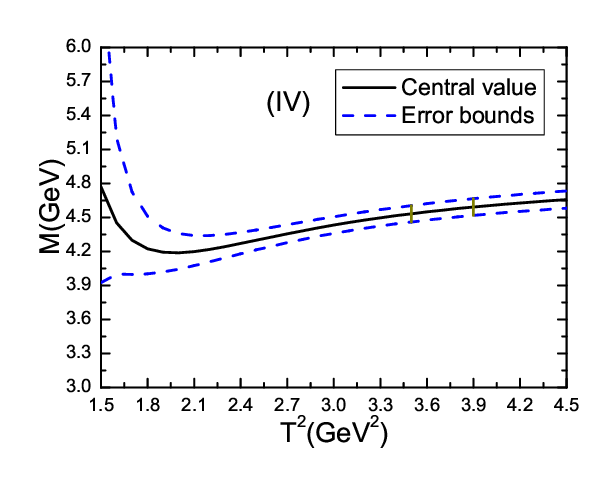}
\includegraphics[totalheight=6cm,width=7cm]{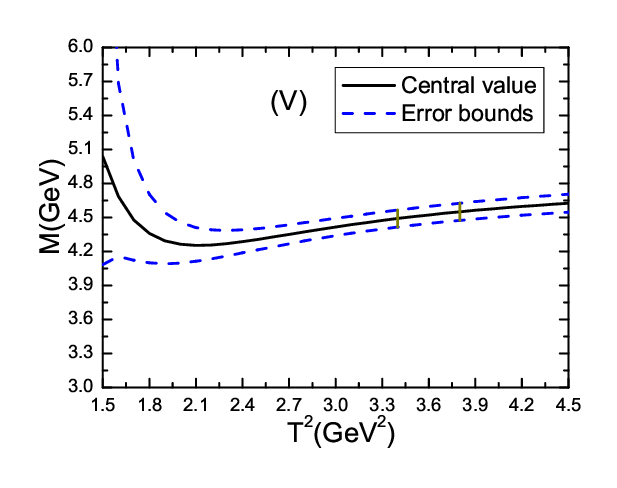}
  \caption{ The masses  with variations of the  Borel parameters $T^2$ for  the hidden-charm-singly-strange pentaquark states, where the (I), (II), (III), (IV) and (V)  represent the
  $[su][uc]\bar{c}$ ($0$, $1$, $1$, $\frac{5}{2}$), $[uu][sc]\bar{c}-[us][uc]\bar{c}$ ($1$, $0$, $1$, $\frac{5}{2}$),  $[uu][sc]\bar{c}-[us][uc]\bar{c}$ ($1$, $1$, $2$, $\frac{5}{2}$), $[su][uc]\bar{c}$ ($1$, $0$, $1$, $\frac{5}{2}$) and $[uu][sc]\bar{c}$ ($1$, $1$, $2$, $\frac{5}{2}$)
    pentaquark states, respectively. }\label{mass-3-fig}
\end{figure}

At last, we take  into account  all uncertainties  of the relevant  parameters,
and obtain  the masses and pole residues of
 the   hidden-charm-singly-strange  pentaquark states with spin-parity $J^P={\frac{1}{2}}^-$, ${\frac{3}{2}}^-$, ${\frac{5}{2}}^-$, which are shown clearly in Figs.\ref{mass-1-fig}-\ref{mass-3-fig} and Table \ref{mass-Pcs}. From Tables \ref{Borel}-\ref{mass-Pcs}, we can obtain the conclusion confidently  that the modified energy scale formula
 $\mu =\sqrt{M^2_{P}-(2{\mathbb{M}}_c)^2}-{\mathbb{M}}_s$ is satisfied very good, just like in our previous works \cite{WangZG-Pc12-Jpsip,WangZG-Pc12-JpsiLambda,WangZG-Pc12-JpsiXi}.
 The (modified)  energy scale formula can enhance the ground state (or pole) contributions  noticeably    and improve the convergent behavior of the operator product expansion  noticeably \cite{WangZG-Review,WangZG-IJMPA-3-scheme}, see Table \ref{Borel}, it is the
 unique and outstanding feature of our works.

 In addition, we can obtain the masses and pole residues of the hidden-charm pentaquark states with the positive parity via the QCD sum rules in Eq.\eqref{QCDSR-Positive}. For example, in the case of the current $J^5(x)$, if we choose the parameters $\sqrt{s_0^\prime}=(5.50\pm0.10)\,\rm{GeV}$ and $T^2=(3.4-3.8)\,\rm{GeV}^2$, then we obtain the pole contribution $(41-62)\%$ and contribution of the vacuum condensates of dimension $13$, $D(13)\ll1\%$. At last, we obtain the mass and pole residue $M_{+}=(4.80\pm0.11)\,\rm{GeV}$ and $\lambda_{+}=(2.83\pm0.51)\times 10^{-3}\,\rm{GeV}^2$, respectively. Thus $M_{+}<\sqrt{s_0}$, there exist contaminations if we choose the traditional QCD sum rules in Eqs.\eqref{Traditional-QCDSR-1}-\eqref{Traditional-QCDSR-0} by setting $\lambda_{+}=0$.

In Figs.\ref{mass-1-fig}-\ref{mass-3-fig}, we plot the masses of the hidden-charm-singly-strange pentaquark states with the isospin $I=1$ according to variations of the Borel parameters in very large ranges, i.e. $T^2=1.5\sim 4.5\,\rm{GeV}^2$, where the regions between the two short vertical  lines are the Borel windows. In the Borel windows, there appear very flat platforms indeed, the uncertainties come from the Borel parameters are very small.

From Table \ref{mass-Pcs}, also in our previous works \cite{WangZG-Pc12-Jpsip,WangZG-Pc12-JpsiLambda,WangZG-Pc12-JpsiXi},  we could  obtain the conclusion confidently  that the lowest hidden-charm pentaquark states are not of the scalar-diquark-scalar-diquark-antiquark type, it is not suitable to call  the scalar and axialvector diquarks as the "good" and "bad" diquarks, respectively \cite{Jaffe-PRT-2005},  because  the scalar diquarks are not necessary to be more stable than the axialvector diquarks.

Quark models usually suggest that the scalar diquark is lighter than the
axialvector one.  For example, the one-gluon exchange induced  color-spin
 interaction leads to the
mass difference between "good" and "bad" diquarks  $\sim \frac{2}{3}
(M_{\Delta}-M_{N})\sim 200\, \rm{MeV}$  from the $\Delta$--nucleon mass difference \cite{Jaffe-PRT-2005}.
If we choose the same valence quarks and adopt the predictions  of the QCD sum rules, the $qq^\prime$-type scalar diquarks have slightly larger masses than the corresponding  axialvector diquarks \cite{WangZG-L-diquark-CTP}, while the $qQ$-type scalar and axialvector diquarks have almost degenerated masses \cite{WangZG-HL-diquark-EPJC,ZhangAL-HL-diquark-PRD}. Naively, we expect that the lowest and most stable configurations are of the scalar-diquark-scalar-diquark-antiquark type, not of the axialvector-diquark-axialvector-diquark-antiquark type.
We should bear in mind that the physical hadron masses are not simple summaries of the masses of all the constituents, but result from complex dynamics of the full QCD. The axialvector diquarks also serve as stable configurations in building the hidden-charm pentaquark states \cite{WangZG-Review}.

We can take the pole residues as basic  input parameters and study the two-body strong decays,
 \begin{eqnarray}
P^+_{cs}&\to& \bar{D}_s\Sigma_c(4423)\, , \,\bar{D}_s\Sigma^*_c(4487)\, , \,\bar{D}^*_s\Sigma_c(4566)\, , \, \bar{D}^*_s\Sigma^*_c(4630)\, , \,\bar{D}\Xi_c(4333)\, ,  \nonumber\\
&&\,\bar{D}^*\Xi_c(4475)\, , \,J/\psi \Sigma(4286) \, , \, \eta_c \Sigma(4173) \, ,
\end{eqnarray}
 with the three-point QCD sum rules to estimate the decay widths and choose the
ideal channels to search for those pentaquark states $P_{cs}^+$, where the superscript $+$ stands for the positive electric charge,  we present the thresholds of the meson-baryon pairs  in the unit $\rm{MeV}$ in the bracket. As the two octets ${\mathbf{8}}_1$ and ${\mathbf{8}}_2$ in Eq.\eqref{two-octet} could mix with each other, the decays to the charmonium plus light baryon octet are allowed.

Up to now, we could reproduce the masses of the existing $P_c$ and $P_{cs}$ states   both in the diquark-diquark-antiquark type and color singlet-singlet type hidden-charm pentaquark scenarios except for the $P_c(4337)$, whose mass could  be reproduced only in the diquark-diquark-antiquark type pentaquark scenario \cite{WangZG-Pc12-Jpsip}. Compared to the color singlet-singlet type pentaquark states, the diquark-diquark-antiquark type pentaquark states have a more copious   spectroscopy. However, we cannot assign a hadron unambiguously  based on the mass alone, furthermore, the masses have uncertainties. We should study their strong decays exclusively and examine the partial decay widths carefully to diagnose their nature as different substructures lead to quite different branching fractions, which are expected to be our next works.

Experimentally, the $P_c^+(4312/4380/4440/4457)$ and $P_{cs}^0(4338/4459)$ were observed in the $J/\psi p$ and $J/\psi \Lambda$ invariant mass distributions, respectively,
\begin{eqnarray}
\Lambda_b^0&\to& P_{c}^+K^- \to J/\psi p \,K^-\, ,\nonumber\\
\Xi_b^-&\to& P_{cs}^0 K^- \to J/\psi \Lambda^0\, K^-\, .
\end{eqnarray}
Accordingly, we expect to observe the  $P_{cs}^+$  states in the $J/\psi \Sigma^+$ invariant mass  distributions  in the weak decays of the ground state bottom baryons,
\begin{eqnarray}
\Sigma_b^+&\to& P_{cs}^+\phi \to J/\psi \Sigma^+ \,\phi\, ,\nonumber\\
\Xi_b^0&\to& P_{cs}^+ K^- \to J/\psi \Sigma^+\, K^-\, ,
\end{eqnarray}
through the CKM favored process $b \to c\bar{c}s$ at the quark level, as the diquark-diquark-antiquark type  and color singlet-singlet  type pentaquark states have quite different mass spectroscopies and testing the predictions is of crucial importance to reach reasonable  conclusion preliminarily  \cite{Review-mole-GuoFK-CTP,WangZG-Review,
Pcs4338-mole-XWWang,Pcs4459-mole-WangZG-SR,Pc4312-mole-penta-WXW-SCPMA,
Pc4312-mole-penta-WXW-IJMPA}. The observations of the $P_{cs}^+$ states in the $J/\psi \Sigma^+$ invariant mass distribution, would deepen our understanding the nature of the $P_{cs}^0(4338/4459)$ in the $J/\psi \Lambda^0$ invariant mass distribution according to conservation of the isospin in the strong decays, and the $\Sigma$ and $\Lambda$ baryons are isospin cousins.

\section{Conclusion}
 In this  work, we adopt the diquark model, construct the diquark-diquark-antiquark type  currents with the light quarks $uus$ in two  octets, and  study  the hidden-charm-singly-strange pentaquark states in the framework of   the QCD sum rules systematically. Routinely, we carry out the operator product expansion to  the    vacuum condensates up to dimension $13$ consistently in our unique scheme, obtain the QCD spectral densities and distinguish  the contributions from the
  $P_{cs}^+$ states with the negative parity unambiguously,
 then we adopt   the modified energy scale formula $\mu=\sqrt{M_{P}-(2{\mathbb{M}}_c)^2}-{\mathbb{M}}_s$ to choose  the optimal  energy scales. Finally, we obtain the mass spectrum (and pole residues as a byproduct) of the hidden-charm-singly-strange pentaquark states with the isospin-spin-parity $IJ^{P}=1{\frac{1}{2}}^-$, $1{\frac{3}{2}}^-$, $1{\frac{5}{2}}^-$, which can be compared with the experimental data in the future, in particular  the processes  $\Sigma_b^+\to P_{cs}^+\phi \to J/\psi \Sigma^+ \,\phi$
and $\Xi_b^0\to P_{cs}^+ K^- \to J/\psi \Sigma^+\, K^-$. As there exist several interpretations for the pentaquark candidates $P_c^+(4312/4337/4380/4440/4457)$ and $P_{cs}^0(4338/4459)$, the observation of the $P_{cs}^+(uusc\bar{c})$ states is of great important to diagnose their nature and distinguish the pentaquark and molecule  interpretations.

\section*{Acknowledgements}
This  work is supported by National Natural Science Foundation, Grant Number  12575083.

\end{document}